\documentclass{article}

\usepackage{arxiv}

\usepackage{amsmath}
\usepackage[utf8]{inputenc} 	
\usepackage[T1]{fontenc}    	
\usepackage{hyperref}       	
\usepackage{url}            	
\usepackage{booktabs}       	
\usepackage{amsfonts}       	
\usepackage{nicefrac}       	
\usepackage{microtype}      	
\usepackage{lipsum}
\usepackage{graphicx}
\usepackage{wasysym}
\usepackage{enumitem}
\usepackage{soul,color}
\usepackage[labelfont=bf]{caption}
\usepackage{setspace}
\usepackage{lineno}
\usepackage{xcolor}
\usepackage[normalem]{ulem}
\usepackage{threeparttable}
\usepackage{siunitx}
\usepackage{multirow}

\graphicspath{ {./figures/} }	

\title{Multi-contrast laser endoscopy for in vivo gastrointestinal imaging}

\author{
Taylor L. Bobrow\\
Department of Biomedical Engineering\\
Johns Hopkins University\\
Baltimore, MD 21218\\
\texttt{tbobrow1@jhmi.edu}\\
\And
Mayank Golhar\\
Department of Biomedical Engineering\\
Johns Hopkins University\\
Baltimore, MD 21218\\
\texttt{m.golhar@jhmi.edu}\\
\And
Suchapa Arayakarnkul\\
Division of Gastroenterology and Hepatology\\
Johns Hopkins Hospital\\
Baltimore, MD 21287\\
\texttt{suchapa.ara@gmail.com}\\
\And
Anthony A. Song\\
Department of Biomedical Engineering\\
Johns Hopkins University\\
Baltimore, MD 21218\\
\texttt{asong18@jhu.edu}\\
\And
Saowanee Ngamruengphong\\
Division of Gastroenterology and Hepatology\\
Johns Hopkins Hospital\\
Baltimore, MD 21287\\
\texttt{sngamru1@jhmi.edu}\\
\And
Nicholas J. Durr\textsuperscript{*}\\
Department of Biomedical Engineering\\
Johns Hopkins University\\
Baltimore, MD 21218\\
\texttt{ndurr@jhu.edu}\\
}

\begin{document}

\maketitle


\begin{abstract}
White light endoscopy is the clinical gold standard for detecting diseases in the gastrointestinal tract. Most applications involve identifying visual abnormalities in tissue color, texture, and shape. Unfortunately, the contrast of these features is often subtle, causing many clinically relevant cases to go undetected. To overcome this challenge, we introduce Multi-contrast Laser Endoscopy (MLE): a platform for widefield clinical imaging with rapidly tunable spectral, coherent, and directional illumination. We demonstrate three capabilities of MLE: enhancing tissue chromophore contrast with multispectral diffuse reflectance, quantifying blood flow using laser speckle contrast imaging, and characterizing mucosal topography using photometric stereo. We validate MLE with benchtop models, then demonstrate MLE \textit{in~vivo} during clinical colonoscopies. MLE images from 31 polyps demonstrate an approximate three-fold improvement in contrast and a five-fold improvement in color difference compared to white light and narrow band imaging. With the ability to reveal multiple complementary types of tissue contrast while seamlessly integrating into the clinical environment, MLE shows promise as an investigative tool to improve gastrointestinal imaging.
\end{abstract}

\raggedbottom

\clearpage

\section*{Introduction}

Flexible endoscopy is a foundational tool for the screening, treatment, and longitudinal assessment of many gastrointestinal disorders, including inflammation, early-stage dysplasia, and cancer. The standard of care is high-definition~(HD) white light endoscopy~(WLE), which acquires high-resolution (\textasciitilde1~megapixel) widefield images with a red-green-blue~(RGB) camera sensor and a broadband illumination source. Gastroenterologists examine endoscopic videos in real time for differences in mucosal color, texture, and shape to identify regions of tissue with abnormalities. Unfortunately, the differences between normal and diseased tissue are often when visualized with WLE. For example, in colonoscopy screening for colorectal cancer (CRC), precancerous adenomas are frequently missed (26\%)~\cite{zhao2019}, reducing its protective value (\textless50\% reduction in CRC mortality)~\cite{singh2010,baxter2009,ransohoff2009}. In the upper gastrointestinal tract, Barrett’s esophagus presents similar diagnostic challenges, as dysplasia often presents with subtle, heterogeneously distributed color changes that can be difficult to detect endoscopically~\cite{spechler2011}. As a result, surveillance of Barrett’s esophagus relies on random four-quadrant biopsies~\cite{qumseya2019} that sample only a small fraction of the at-risk mucosa, resulting in an estimated 25\% miss rate for esophageal adenocarcinomas~\cite{sawas2022}.

The limited contrast of WLE has motivated the development of complementary optical imaging modalities to improve the detection of abnormal tissues. For example, narrow band imaging (NBI) illuminates tissue with narrow spectral bands of light centered at 415~nm and 540~nm~\cite{song2008}. These colors correspond to spectral absorption peaks of hemoglobin and are rapidly attenuated in tissues, resulting in the enhancement of vascular contrast and the reduction of deep-tissue background signal. Clinical evidence suggests that NBI is an effective tool for predicting lesion histology~\cite{mcgill2013} and improving targeted biopsy sampling of Barrett’s esophagus~\cite{qumseya2013}. However, it provides little to no improvement in adenoma detection rates during colonoscopy (\textless3\%)~\cite{dinesen2012,atkinson2019}.

Beyond vascular contrast, other technologies have explored leveraging cellular-level structural contrast between normal and diseased tissues, including elastic scattering spectroscopy~\cite{mourant1996}, confocal endomicroscopy~\cite{kiesslich2004,xie2011}, and optical coherence tomography~\cite{tearney1996} in both the lower~\cite{dhar2006,rodriguez2014,hurlstone2008,jackle2000,nie2024} and upper gastrointestinal tracts~\cite{jackle2000,lovat2006,kiesslich2006}. Although these modalities are promising for lesion classification and margin delineation, their small field-of-view~(FoV) inhibits the surveillance of large areas of the gastrointestinal tract, making them impractical for lesion detection. In addition to endogenous sources of contrast, exogenous fluorescent dyes have been designed to selectively bind to tumor-associated molecules for the detection and delineation of lesions~\cite{garcia2013,yim2021,thekkek2013}. The specificity, cost, and challenges associated with dye administration remain significant obstacles to clinical adoption.

Both technical challenges associated with imaging through flexible endoscopes, as well as safety barriers associated with human testing, have hindered the \textit{in~vivo} exploration of novel widefield optical enhancement modalities in gastroenterology. To circumvent these challenges, researchers often characterize resected, \textit{ex~vivo} tissue as a surrogate for \textit{in~vivo} tissue. Unfortunately, resected tissues undergo rapid chemical and morphological changes, introducing significant error to the measured biomarkers~\cite{waterhouse2019,vaupel2021}. To enable \textit{in~vivo} imaging, researchers have developed probes and imaging bundles that can be inserted through the endoscope instrument channel~\cite{harmsen2019,kester2011,yoon2019}. A major technical challenge of this approach is the limited spatial resolution of the imaging bundles, which reduces the image resolution by several orders of magnitude compared to conventional flexible endoscopes. Moreover, the utilization of the instrument channel for imaging requires frequent insertion and removal of the probe, significantly lengthening procedures. Another approach is to modify the clinical illumination source to facilitate the transmission of light through the endoscope's fiber optic light guide~\cite{han2016}. Although this approach significantly improves image resolution, it also eliminates the ability to image with clinical illumination, preventing the delivery of the clinical standard-of-care. To improve cancer detection in endoscopy, enhanced imaging techniques must be capable of non-contact visualization of a wide field-of-view at varying working distances, with HD resolution, and at a video frame rate. 

Here we introduce Multi-contrast Laser Endoscopy (MLE) - a research platform for acquiring multimodal images of gastrointestinal tissue \textit{in~vivo}. This approach retrofits the internal light guides of a clinical colonoscope, incorporates a multi-contrast laser illumination source, and utilizes a custom hardware-software architecture for synchronized illumination control and real-time frame capture, processing, and display. The MLE system can display conventional standard-of-care WLE images while acquiring multimodal data in the background, facilitating its integration into the clinical workflows. We validate multispectral, perfusion, and topographic MLE modalities using calibrated imaging targets and phantom tissue models. MLE’s capabilities are then demonstrated in a first-in-human pilot study of participants undergoing standard-of-care colonoscopy. Our results demonstrate that MLE provides real-time, high-definition, wide field-of-view contrast enhancement of pathological gastrointestinal tissues \textit{in~vivo}.

\section*{Results}

\subsection*{Design of a clinically translatable MLE system}

\begin{figure}
\centering
\includegraphics{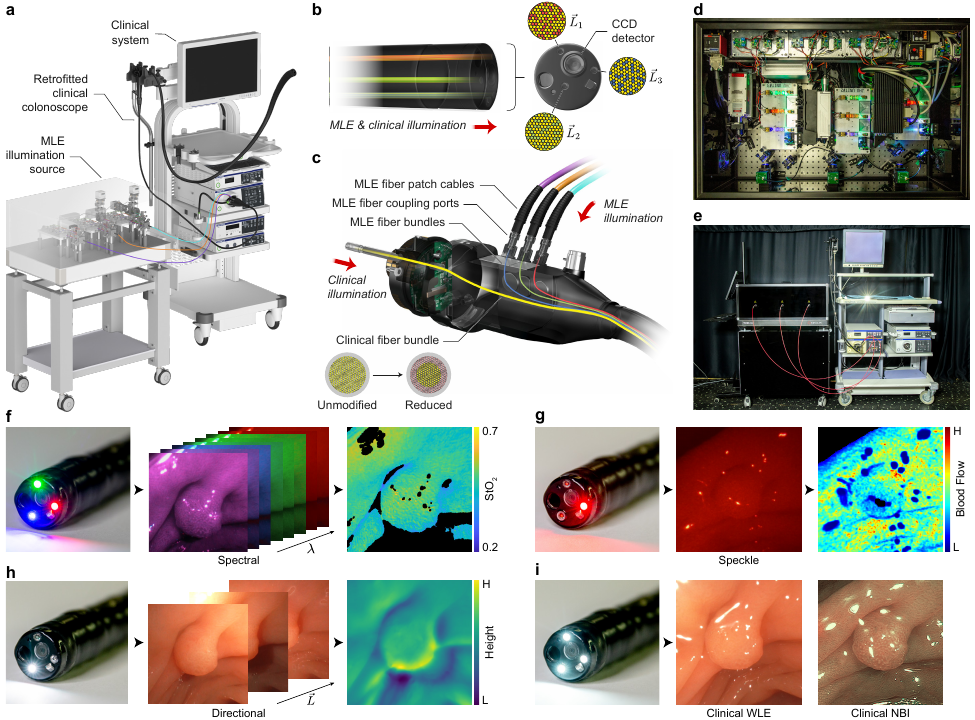}
\caption{\textbf{The Multi-contrast Laser Endoscopy (MLE) system for in~vivo gastrointestinal imaging.} \textbf{a} MLE is built around a retrofitted clinical colonoscope that interfaces with a standard clinical endoscopic imaging system while accepting multi-contrast laser illumination. \textbf{b,~c} The internal light guide of a clinical colonoscope was replaced with a custom fiber optic bundle for transmitting both clinical and MLE illumination. MLE illumination is coupled through three fiber ports installed in the connector-end of the scope (c). Both the clinical fiber bundle and MLE fiber bundles map to three point sources surrounding the CCD detector at the distal tip of the scope (b). \textbf{d} Photograph of the multi-contrast illumination source constructed from lasers with varying wavelengths and coherence lengths. \textbf{e} Photograph of the retrofit clinical colonoscope, clinical endoscopic imaging system, and multi-contrast laser illumination source. \textbf{f~-~i} Tubular adenoma (5~mm diameter) in the descending colon imaged during a screening colonoscopy. \textbf{f} Multispectral frames were acquired with illumination from 8 narrow-band wavelengths ($\lambda$) and used to estimate total oxygen saturation. \textbf{g} High-coherence illumination generates laser speckle contrast for visualizing local tissue perfusion. \textbf{h} Point sources surrounding the detector ($\vec{L}$) were independently toggled to acquire directionally illuminated images to enhance tissue surface shape. \textbf{i} The system retains full standard-of-care clinical imaging capabilities for acquiring white light endoscopy (WLE) and narrow band imaging (NBI) illuminated frames.}
\label{fig:Figure1}
\end{figure}

To enable high-throughput experimental imaging of gastrointestinal tissues in the clinical setting, we retrofitted a clinical colonoscope to accept a custom-built multi-laser illumination source and interface with standard clinical endoscopic imaging systems (Fig.~\ref{fig:Figure1}a). Unlike previous systems that use imaging fiber bundles for data acquisition, our system leverages the optimized optics and CCD sensor within the endoscope to acquire HD (1080~\texttimes{}~1350~pixels), video rate (29.97 interlaced frames per second), wide FoV (170\textdegree), large depth-of-field (DoF; 5-100~mm) image data. In addition, the retrofitted colonoscope design enables rapid toggling (\textless1~s) between clinical and MLE illumination modes, minimizing disruption to the clinical workflow.

To facilitate data acquisition with both clinical WLE and MLE illumination modes, a colonoscope was internally modified  with a custom fiber optic light guide (Fig.~\ref{fig:Figure1}b-c). The light guide includes three MLE fiber bundles that are accessible for optical coupling through three ports added to the connector-end of the colonoscope (Fig.~\ref{fig:Figure1}c). Fibers from the MLE bundles are randomly mixed with fibers from the clinical illumination bundle at the distal end of the colonoscope (Fig.~\ref{fig:Figure1}b), allowing tissue to be imaged with light from either illumination source. This approach reduces the barrier to clinical translation, as the system maintains full clinical functionality for delivering the standard-of-care, and modifications are limited to non-patient contacting surfaces of the colonoscope, minimizing any added risk of infection or other complications.

The multi-laser illumination source was constructed using 15 laser diodes with 8 narrowband wavelengths (406-657~nm) and varying coherence lengths to acquire speckle and speckle-free images (Fig.~\ref{fig:Figure1}d). A light modulation controller was developed to synchronize the pulse widths of the laser diodes with the frame capture rate of the colonoscope CCD sensor, and a custom software platform was developed to acquire, display, and store raw uncompressed frames from the clinical endoscopy system in real time. The system includes an auto-exposure function to prevent over- and under-exposed frames during the frequent changes in working distance typical in endoscopy. The illumination source was built on an optical breadboard enclosed in a cart for transport to and from the clinic (Fig.~\ref{fig:Figure1}e). The MLE system can rapidly acquire multiple contrast enhancement channels, including spectral, flow, and topographic measurements, in addition to clinical WLE and NBI (Fig.~\ref{fig:Figure1}f-i).

\subsection*{Widefield measurement of spectral reflectance and oxygen saturation}

\begin{figure}
\centering
\includegraphics{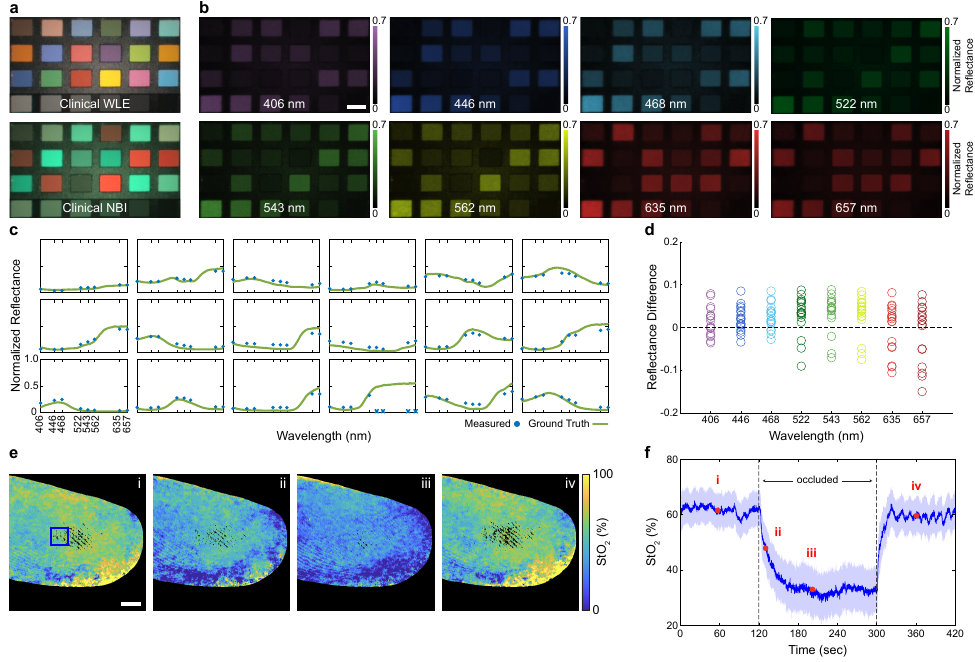}
\caption{\textbf{MLE enables widefield spectral imaging for interrogating molecular chromophore changes in tissue.} \textbf{a} Images of a Macbeth color target acquired with clinical WLE and NBI illumination. \textbf{b} Multispectral reflectance images of the color target acquired with MLE. Reflectance measurements were normalized using images of a white reference target. \textbf{c} Mean spectral reflectance values from the first three rows of patches in (b) and ground truth spectra measured with a spectrometer. Four measurements with saturated pixels are indicated by an X. \textbf{d} Relative difference between MLE reflectance and reference spectrometer measurements for each wavelength channel. \textbf{e} Total oxygen saturation ($\mathrm{StO_2}$) measurements of a human index finger before (i), during (ii,iii), and after occlusion (iv) by a pressure cuff. Black regions indicate saturated, under-exposed, and background pixels. \textbf{f} Time series $\mathrm{StO_2}$ values from the region indicated by the blue square in (e). Solid line and shaded area in (f) indicate the $\mathrm{StO_2}$ mean and standard deviation, respectively. See Supplementary Movie 1 for measurements over the complete time course. Scale bars~=~4~mm}
\label{fig:Figure2}
\end{figure}

Diffuse reflectance spectroscopy has been widely studied as a non-contact, label-free method for revealing differences in underlying tissue makeup and structure. To evaluate the accuracy of multispectral reflectance measurements acquired with MLE, we imaged a color calibration chart with 18 patches that exhibit distinct spectral reflectance signatures. Compared to clinical WLE and NBI, which capture two to three bands of spectral information (Fig.~\ref{fig:Figure2}a), MLE measures spectral reflectance at 8 distinct wavelengths (Fig.~\ref{fig:Figure2}b). Qualitative inspection of the spectral reflectance values measured by MLE and a reference spectrometer shows strong agreement for each Macbeth patch (Fig.~\ref{fig:Figure2}c). Quantitative comparison with the reference spectra results in a mean absolute reflectance error of 0.04~$\pm$~0.03, with each spectral channel error centered near 0 (Fig.~\ref{fig:Figure2}d).

Spectral information can be used to map absorbance and quantify hemoglobin concentration and total oxygen saturation ($\mathrm{StO_2}$) levels, both of which have been observed to change with cancer progression~\cite{muz2015}. To assess the ability of MLE to detect changes in these biomarkers, we conducted a finger occlusion trial where tissue ischemia was systematically induced by inflating a pressure cuff. The tip of the colonoscope was placed above the index finger, while reflectance measurements were acquired cyclically at a rate of 7.5 full spectral measurements per second. The reflectance images were converted to absorption (see Methods), and the relative concentrations of oxy- and deoxy-hemoglobin were estimated by fitting the absorption data to known reference spectra. Sample $\mathrm{StO_2}$ maps from before (i), during (ii,~iii), and after (iv) occlusion are shown in (Fig.~\ref{fig:Figure2}e). $\mathrm{StO_2}$ values from a 50~\texttimes{}~50~pixel region of interest during the full-time course are shown in Fig.~\ref{fig:Figure2}f. The measured oxygen saturation dropped from a baseline of 61.5~$\pm$~5.9\% (i) to 33.0~$\pm$~8.0\% during occlusion (iii), before returning to 59.6~$\pm$~6.0\% (iv) after releasing the pressure cuff. The oxygen saturation measurements acquired with MLE are consistent with measurements acquired with a benchtop imaging system~\cite{chen2020} and demonstrate the sensitivity of MLE to measuring functional differences in tissue.

\subsection*{Laser speckle contrast enhancement of flow regions}

\begin{figure}
\centering
\includegraphics{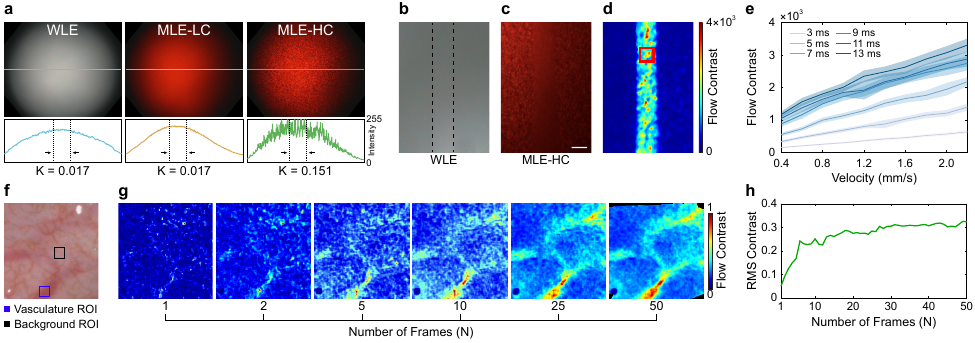}
\caption{\textbf{MLE is sensitive to differences in blood flow.} \textbf{a} Images of a uniform scattering medium under WLE, low coherence laser (MLE-LC), and high coherence laser (MLE-HC) illumination. Displayed below are line profiles from the red color channel of each image, along with speckle contrast (K) values taken from $\pm$100~pixels around the center of the full-width half maximum of each profile. Low-coherence laser illumination provides similar illumination uniformity and smoothness to WLE. \textbf{b,~c} Images of a microfluidic phantom under WLE (b) and MLE-HC (c) illumination. A polystyrene bead suspension was pumped through a microfluidic phantom channel (dashed lines in b) at a velocity of 1.1~mm/s to simulate blood flow.  \textbf{d} Flow contrast map generated by applying laser speckle contrast analysis to (b). A 9~\texttimes{}~9 median filter was applied for visualization. \textbf{e} Flow contrast values from the region indicated by the red square in (d) across a range of flow velocities and laser pulse widths. Solid lines and shaded areas indicate the flow contrast mean and standard deviation from 100 flow contrast images, respectively. \textbf{f} WLE image of arterioles and capillaries in the soft palate. \textbf{g} Flow contrast maps of the field of view (FoV) shown in (f) with an increasing number of registered and averaged flow contrast frames. See Supplementary Movie 2 for a flow contrast video sequence. Black regions indicate either saturated or underexposed pixels, or pixels outside the FoV due to motion registration. \textbf{h} RMS contrast for flow contrast frames between the regions indicated in (f). All values are reported in arbitrary units. Scale bar~=~100~\textmu{}m}
\label{fig:Figure3}
\end{figure}

In addition to the low-coherence laser diodes (MLE-LC) used for spectral imaging, a high-coherence laser (MLE-HC) was integrated into the multi-contrast illumination source to produce well-defined speckle for laser speckle contrast imaging (LSCI). As shown in Fig.~\ref{fig:Figure3}a, MLE-LC illumination produce images with speckle contrast ($\mathrm{K}$) similar to clinical WLE, while MLE-HC illumination produces images with an order of magnitude greater speckle contrast.

To characterize the ability of MLE to quantify flow with LSCI, a microfluidic flow phantom was fabricated from polydimethylsiloxane~(PDMS) doped with titanium dioxide (TiO\textsubscript{2}) and India ink to simulate optical scattering and absorption by tissue. An open 500~\texttimes{}~150~\textmu{}m channel was incorporated 100~\textmu{}m below the phantom surface. A suspension of polystyrene beads was mixed to emulate the optical scattering of human blood, and a syringe pump was used to pump the microsphere solution through the channel at velocities spanning flow rates typically observed in microvasculature (0.4~mm/s~-~2.2~mm/s). Fig.~\ref{fig:Figure3}b,c show images of the phantom acquired with clinical WLE and MLE-HC illumination. In both images, the flow chamber is indistinguishable from the static surface surrounding it. However, after processing the MLE-HC image with LSCI, which spatially analyzes the reduction in speckle intensity caused by motion, the flow region becomes visually identifiable (Fig.~\ref{fig:Figure3}d). In addition to improving the visual contrast of the flow regions, LSCI can quantify differences in flow velocity (Fig~\ref{fig:Figure3}e). The results are consistent with previous studies acquired with non-endoscopic systems, in which the flow contrast increases with flow velocity and exposure time~\cite{thompson2008,kazmi2013}. The slope of contrast versus velocity differed significantly across exposure times (ANCOVA, velocity~\texttimes~exposure interaction: F(5,~48)~=~101.39, $p$~<~0.001; Supplementary~Fig.~6), indicating that longer exposure times amplify the rate at which contrast increases with velocity.

To assess the sensitivity of MLE to variations in blood flow in the presence of motion \textit{in~vivo}, we imaged arterioles and capillaries in the human soft palate mucosa (Fig.~\ref{fig:Figure3}f). The flow contrast was improved by registering sequential speckle contrast frames, via interlaced color imaging, and averaging them (see Methods). Qualitative inspection of the flow contrast maps shows that the vasculature becomes distinguishable from the background tissue with an average of 10 or more frames (Fig.~\ref{fig:Figure3}g). To quantify this improvement, the root-mean-square (RMS) contrast~\cite{qi2023} between the vasculature and background regions of interest (ROI) was calculated for each temporal window size (Fig.~\ref{fig:Figure3}h). These results demonstrate that RMS contrast increases significantly over the first 1-6 frames, then gradually improves as the number of frames increases.

\subsection*{Surface shape enhancement with photometric stereo}

Tissue shape is an important cue for detecting and classifying adenomas~\cite{shandro2020}. Modern colonoscopes illuminate the field through multiple light sources oriented around the camera. While this approach minimizes shadowing and improves illumination quality, it also minimizes topographical shading variations that improve the perception of surface shape. To enable direct measurement of colon topography, we captured directionally illuminated images by sequentially toggling laser sources that map to the three point sources at the tip of the scope ($\vec{L}_1$~-~$\vec{L}_3$). This sequence of images was used to computationally reconstruct surface topography using photometric stereo endoscopy (PSE)~\cite{parot2013}.

\begin{figure}
\centering
\includegraphics{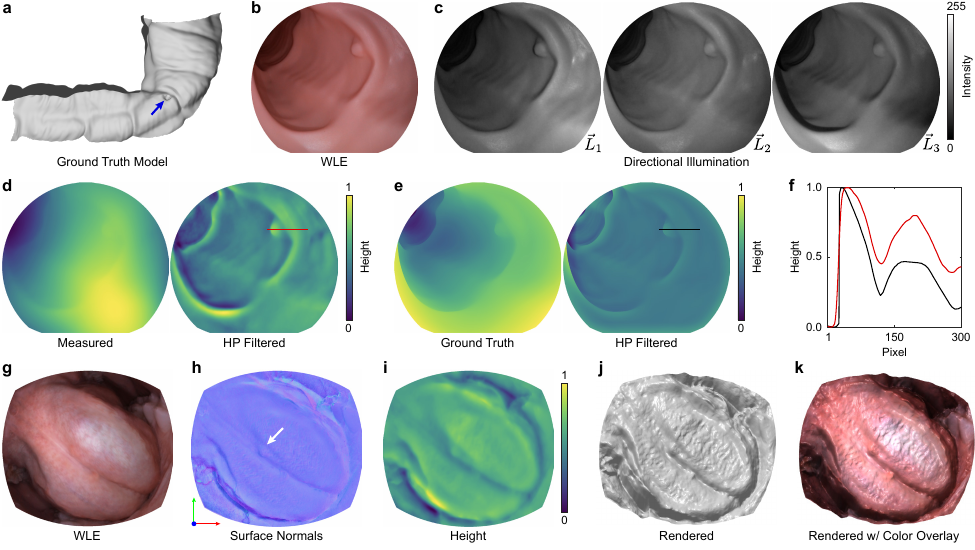}
\caption{\textbf{MLE enhances visual contrast of subtle changes in mucosal surface.} \textbf{a} Virtual 3D colon model with a 7~mm diameter tubular adenoma (blue arrow). \textbf{b}  WLE image of a silicone cast of the model. \textbf{c} MLE images of the phantom acquired with directional illumination from the three point sources ($\vec{L}$) at the tip of the colonoscope ($\vec{L}_1$, upper right; $\vec{L}_2$, upper left; $\vec{L}_3$, lower left). \textbf{d} Surface height maps estimated from the images in (c) before (left) and after (right) applying a high-pass (HP) spatial frequency filter. \textbf{e} Ground truth surface height maps rendered from the virtual 3D colon model before (left) and after (right) applying a HP spatial frequency filter. \textbf{f} Comparison of the normalized surface height values from the red (measured) and black (ground truth) lines shown in (d) and (e). \textbf{g} WLE image of the ventral surface of the tongue. \textbf{h,i} HP filtered surface normal and height maps computed from MLE directionally illuminated images. The white arrow points to a diminutive surface feature that is not visible in the WLE image. \textbf{j,k} Surface height map from (i) rendered using a virtual directional illumination source, a Phong reflection model, with a constant albedo (j) and a color overlay (k). Surface normal maps are displayed with the X/Y/Z components stored in separate R/G/B color channels. The components are linearly scaled from $\pm$1 to 0-255. Height values are normalized. See Supplementary Movie 3 for ventral tongue measurements over the complete time course.}
\label{fig:Figure4}
\end{figure}

We characterized the accuracy of PSE  topography estimation by measuring a silicone colon phantom and comparing surface reconstructions to the known model profile (Fig.~\ref{fig:Figure4}a). We captured images of the phantom with both WLE and directional illumination (Fig.~\ref{fig:Figure4}b,c). We then estimated the surface normal vector of each image pixel using a fixed geometry between the camera and the light sources. This surface map was integrated to produce a surface height map (Fig.~\ref{fig:Figure4}d). Low-spatial frequency error in the height map resulting from the unknown working distance between the endoscope and the tissue surface was attenuated by applying a high-pass filter. The  high-spatial frequencies of the measured height map show strong agreement with the ground truth (Fig.~\ref{fig:Figure4}e), as shown in Fig.~\ref{fig:Figure4}f, achieving a mean absolute error of 0.11.

PSE was then tested in human tissue by free hand imaging the ventral tongue (Fig.~\ref{fig:Figure4}g). The directionally illuminated images were registered and spatially aligned before estimating high-spatial frequency surface normal and height maps (Fig.~\ref{fig:Figure4}h,i), which improves the visual contrast of diminutive topographical features that were not visible in WLE (Fig.~\ref{fig:Figure4}g). The height map was also rendered for qualitative visualization with and without the WLE measurement overlay (Fig.~\ref{fig:Figure4}j,k).

\subsection*{Application of MLE during colonoscopies}

We conducted a clinical study with MLE, composed of 20 participants scheduled for colonoscopy at the Johns Hopkins Hospital. We imaged 31 polyps from these patients with both WLE clinical and MLE research illumination modes (Supplementary Table~3). Only precancerous polyps (tubular, tubulovillous, or serrated histology) confirmed with histopathology of the resected samples were included in the study.

\begin{figure}
\centering
\includegraphics{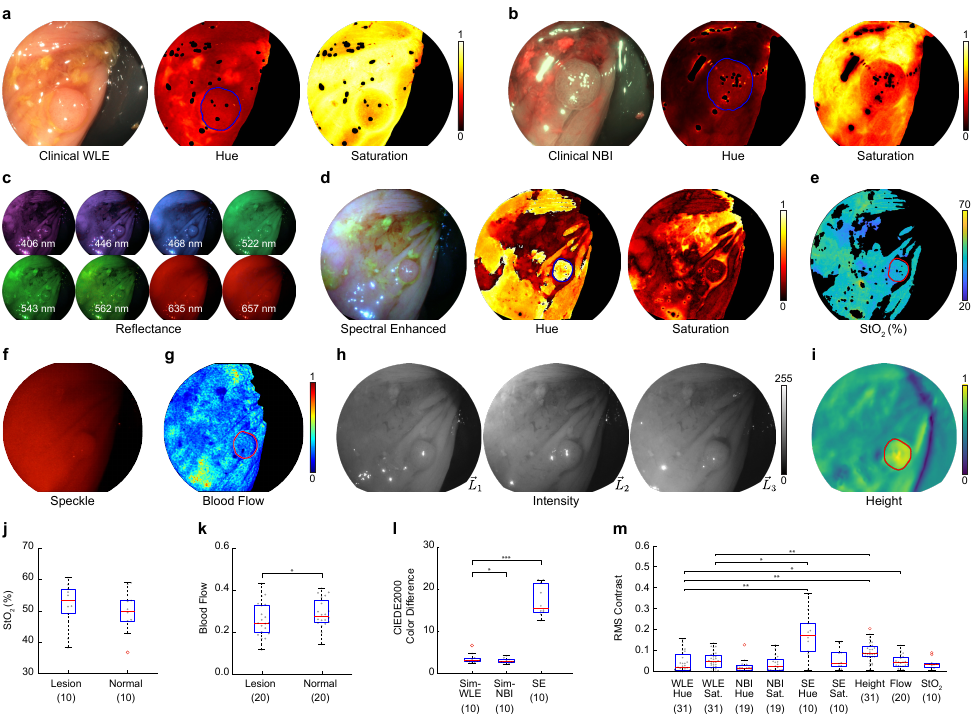}
\caption{\textbf{MLE enhances adenoma contrast during screening colonoscopies.} \textbf{a,~b} Tubular adenoma (4~mm diameter) in the ascending colon visualized with clinical white light endoscopy (WLE) and narrow band imaging (NBI). The WLE and NBI frames were converted to normalized hue and saturation for brightness-independent analysis of color contrast. \textbf{c} Multispectral reflectance images acquired at eight narrowband wavelengths with MLE. \textbf{d} Spectral enhanced image with normalized hue and saturation channels rendered using the reflectance data in (c). \textbf{e} Spatially-mapped total oxygen saturation recovered from the reflectance data in (c). \textbf{f,~g} High-coherence (HC) laser speckle image and spatially-mapped blood flow (normalized). \textbf{h,~i} Directionally illuminated images and recovered height map (normalized). \textbf{j,~k} Average $\mathrm{StO_2}$ (j) and blood flow (k) values from lesion and normal tissue regions in each sample. \textbf{l} CIEDE2000 color difference between pixels from normal and lesion tissue regions in each sample with statistical comparisons made between Sim-WLE and each modality. \textbf{m} RMS contrast values for each modality between normal and lesion tissue regions with statistical comparisons made between WLE-Hue/WLE-Sat. and each modality. Red and blue lines in (a)-(i) denote the boundary between normal and lesion tissue used for analysis. Black regions indicate saturated, under-exposed, and pixels beyond the field of view due to image transformation following registration. See Supplementary Movies 4-6 for videos of measurements over the complete time course. The lines on box and whisker plots show the median, interquartile range, and non-outlier range (outliers are marked as red circles). The number of samples analyzed is listed below each modality in parenthesis, and an outline of samples is provided in Supplementary Table~3. Statistical values are reported in Supplementary Tables~4-6. Significance levels: *~$p$~<~0.05; **~$p$~<~0.01; ***~$p$~<~0.001.}
\label{fig:Figure5}
\end{figure}

An example lesion from the dataset (4~mm tubular adenoma, ascending colon) is shown in Fig.~\ref{fig:Figure5}. WLE visualization exhibits poor contrast between the adenomatous tissue and normal tissue. Both regions appear yellow with subtle perceivable differences in texture or surface shape (Fig.~\ref{fig:Figure5}a). Similarly, under NBI illumination, the color contrast is poor, as both types of tissue have a blue tint (Fig.~\ref{fig:Figure5}b). However, NBI offers some improvement in visualizing the disrupted surface texture compared to WLE, likely because of the shallow penetration depth of the illumination wavelengths. A lack of color contrast is also evident in the hue and saturation channels of both WLE and NBI images.

Using spectral reflectance data acquired with MLE (Fig.~\ref{fig:Figure5}c), we rendered color images while varying the camera's spectral transmission curves to optimize the color contrast between normal and adenomatous tissues (Supplementary Fig.~8). The color contrast in the spectral enhanced (SE) image was significantly improved compared to clinical WLE and NBI, with the lesion appearing darker and red relative to the surrounding tissue (Fig.~\ref{fig:Figure5}d). The improvement in color contrast can also be seen in the hue and saturation channels. Quantitatively, the mean CIEDE2000 color difference increased from 3.1~$\pm$~1.3 for Sim-WLE to 16.7~$\pm$~3.7 for SE, corresponding to a five-fold increase in color contrast. This increase was statistically significant in paired comparisons ($\Delta\mu$~=~13.62, $p$~=~1.04\texttimes10\textsuperscript{-7}, $d_z$~=~4.79; Fig.~\ref{fig:Figure5}l; Supplementary~Table~5). For paired comparisons of RMS contrast with SE-Hue, the differences were statistically significant for WLE-Hue ($\Delta\mu$~=~0.159, $d_z$~=~1.47, $p$~=~0.008) and WLE-Sat. ($\Delta\mu$~=~0.139, $d_z$~=~1.28, $p$~=~0.017; Fig.~\ref{fig:Figure5}m; Supplementary~Table~6).

We also explored using functional measurements as a source of image contrast. The reflectance images from MLE were used to spatially map total oxygen saturation (Fig.~\ref{fig:Figure5}e). The average total oxygen saturation ($\overline{x}$~=~50.1\%$~\pm$~6.9\%) is consistent with prior work~\cite{taylor2024}, but the differences between polyp and normal tissue were not statistically significant ($\Delta\mu$~=~2.1\%, $d_z$~=~0.65, $p$~=~0.07; Fig.~\ref{fig:Figure5}j; Supplementary~Table~4). Laser speckle illuminated frames (Fig.~\ref{fig:Figure5}f) were acquired to spatially map blood flow with LSCI (Fig.~\ref{fig:Figure5}g). The mean blood flow measured with LSCI was less in polyp tissue than in normal tissue and the difference was found to be statistically significant ($\Delta\mu$=-0.03, $d_z$=-0.48, $p$=0.046; Fig.~\ref{fig:Figure5}k; Supplementary~Table~4). In some samples, the decrease in blood flow was relatively uniform within the margins of the polyps, making them visually identifiable, whereas in other cases, the blood flow was non-uniform in appearance. The poor distinction between tissue types is reflected in the quantitative contrast analysis, where RMS contrast values were largely comparable to WLE. Flow showed a statistically significant but small increase relative to WLE-Hue ($\Delta\mu$~=~0.025, $d_z$~=~0.71, $p$~=~0.025), whereas no significant difference was observed relative to WLE-Saturation ($p$~=~0.969).

In contrast to WLE and NBI, height reconstruction from MLE directional illumination (Fig.~\ref{fig:Figure5}h) revealed a distinct change in surface topography at the lesion margins (Fig.~\ref{fig:Figure5}i), offering a valuable signal for lesion detection and segmentation. The RMS contrast provided by surface shape increased from 0.04~$\pm$~0.05 and 0.05~$\pm$~0.04 for WLE-Hue and WLE-Sat. to 0.09~$\pm$~0.05 for height reconstruction, corresponding to an approximately two-fold improvement in contrast. This increase was statistically significant in paired comparisons with WLE-Hue ($\Delta\mu$~=~0.049, $d_z$~=~0.63, $p$~=~0.009) and WLE-Sat. ($\Delta\mu$~=~0.040, $d_z$~=~0.67, $p$~=~0.006; Fig.~\ref{fig:Figure5}m).

\section*{Discussion}

While colonoscopy has long been considered the gold standard for preventing and detecting CRC, recent evidence suggests that its efficacy in reducing CRC mortality may be less robust than initially believed. This efficacy is limited in part from missed lesions that arise from reliance on color and texture information to differentiate diseased and healthy tissue. To improve lesion contrast contrast, optical image enhancement modalities are being developed to offer label-free, real-time information that complements WLE. Here, we introduced an MLE platform to investigate new sources of optical contrast during \textit{in~vivo} gastrointestinal imaging. Unlike other techniques that use custom scopes or probes, this approach integrates directly with the clinical imaging system and leverages the optimized internal optics and image sensor. The result is a system with a spatial resolution (1.5M~pixels), FoV (170\textdegree), and DoF (5-100~mm) equivalent to the standard-of-care. Importantly, the system can quickly toggle between clinical and research illumination modes, allowing its practical use in large clinical studies.

The methods and results presented here come with several important limitations. First, regarding the MLE system, pixel values captured by the colonoscope’s CCD sensor are subject to non-linear and black-box transformations (e.g. gamma correction) by the clinical video processor. Although benchtop characterization with a color target demonstrated robust measurement accuracy, further performance gains could likely be achieved by inverting or circumventing these transformations. Additionally, frames are sequentially acquired with different illumination parameters, and spatially aligned using image registration during post-processing. The frequent movement of the scope and surrounding tissue limits the number of frames, and consequently, the number of illumination channels that can be reliably registered. It is also important to note that, although our results demonstrate that MLE improves the RMS contrast between lesions and surrounding tissue, our study design did not test if this approach translates into an improvement in lesion detection rates. The impact on lesion detection merits further investigation, possibly through a randomized controlled trial that compares the adenoma detection rate (ADR) between WLE and MLE groups.

With simple modifications, the MLE research platform can be utilized to address several key opportunities in endoscopy screening. First, the laser diodes within the illumination source can be interchanged to evaluate different wavelengths for lesion endogenous or exogenous contrast. These wavelengths could be extended to the ultraviolet range for autofluorescence imaging or to the near-infrared range for evaluating deep and background-free fluorescence imaging with molecularly-targeted dyes~\cite{yim2021}. Additionally, the high-coherence channel can be used in conjunction with speckle-illumination spatial frequency domain imaging to quantify bulk tissue optical properties~\cite{chen2021,song2025}. Second, the feature-rich information acquired by MLE could improve the performance of deep learning-based detection, classification, and margin delineation. Third, a learned-sensing approach could enable the joint optimization of illumination parameters (wavelength, coherence, direction) with a neural network's filter weights, thereby optimizing the measurement system for machine interpretation~\cite{muthumbi2019}. The learned illumination weightings could then be used to acquire computational images in a single snapshot, eliminating the need to register frames. Further, the computational image could be digitally transformed to a WLE image for interpretation by the clinician~\cite{bobrow2019}. Learning-based methods require large volumes of data, and MLE addresses this challenge with its relative ease of obtaining large quantities of \textit{in~vivo} data. Finally, the system could enhance esophageal and gastric cancer screening and support applications beyond lesion detection, including quantitative fibrosis grading in IBD and detecting Barrett’s esophagus.

\section*{Methods}

\subsection*{Retrofitted clinical colonoscope design}

The MLE system was constructed around an adult video colonoscope (CF-HQ190L; Olympus America Inc., Center Valley, PA, US) modified to enable illumination with clinical and research illumination sources (Fig.~\ref{fig:Figure1}). The colonoscope’s internal fiber optic light guide was replaced with a custom fiber optic bundle that includes multiple MLE fiber bundles that allow external illumination. The custom bundle was constructed from 35~\textmu{}m, 0.55~NA borosilicate fibers with custom-machined ferrules compatible with the internal components of the scope (Gulf Fiberoptics, Oldsmar, FL, US). To make the internal MLE fiber bundles accessible for optical coupling, the fiber bundle ferrules were mounted within stainless steel SMA fiber ports added to the connector-end of the colonoscope. The fiber port additions and custom fiber bundle installation were made by Fibertech Medical (Timonium, MD, US). See Supplementary Section~1 and Supplementary Fig.~1 for additional fabrication details.

\subsection*{Multi-contrast illumination source}

The multi-contrast illumination source included three RGB laser units to emulate standard white light (enabling all three RGB units) and directional (enabling a single RGB unit) illumination. The RGB units were constructed from 446~nm, 522~nm, and 635~nm multimode laser diodes with collimation and combining optics. To increase the diversity of wavelengths available for multispectral imaging, five additional multimode laser diodes with wavelengths centered at 406~nm, 468~nm, 543~nm, 562~nm, and 657~nm were included in the system. These wavelengths were selected to coincide with points at which the absolute values of the ratios of oxy- and deoxy-hemoglobin are either maximal or close to one (isosbestic) to facilitate tissue oxygen saturation mapping. A 639~nm single longitudinal mode laser (coherence length~>~40~m) was included to illuminate with laser speckle for flow contrast imaging. The laser diodes were powered by drivers equipped with transistor–transistor logic switching for pulse width modulation. The drive current for each laser diode was tuned so that the total output power emitted by the scope with all diode channels enabled was less than the maximum output power emitted by an unmodified colonoscope and clinical light source (330~mW). Laser diode temperature was regulated by PID-controlled thermoelectric cooler drivers. Laser diode output power was continually measured using photodiode monitoring units. See Supplementary Section~2, Supplementary Figs.~2 and 3, and Supplementary Tables~1 and 2 for optical components and layout.

\subsection*{Light modulation controller}

Laser diodes in the multi-contrast illumination source were triggered using a custom light modulation controller printed circuit board (PCB). The PCB contained a video-sync separator chip (LM1881; Texas Instruments, Dallas, TX, US) for synchronizing the light triggering with the colonoscope frame acquisition rate. The PCB also included a microcontroller unit for controlling the output power of each diode with pulse width modulation. See Supplementary Section~3 and Supplementary Fig.~4 for additional design details and a circuit schematic.

\subsection*{System architecture}

The clinical endoscopic imaging system consists of a video processor (CV-190; Olympus America Inc., Center Valley, PA, US), xenon light source (CLV-190; Olympus America Inc., Center Valley, PA, US), and two display monitors. The connector end of the colonoscope plugs into the clinical light source and video processor for optical coupling with the arc lamp and transmission of the CCD signal to the video processor. Control of the MLE system was managed by a workstation (Dell Precision 5820T, Intel i7-9800X 3.80~GHz 8-core Central Processing Unit (CPU); 64.0~Gb physical memory) running Microsoft Windows~10. An HD frame grabber (Orion HD, Matrox Imaging, Montreal, Canada) captured and stored uncompressed video from the clinical video processor. A custom C++ application was developed to enable real-time data acquisition, control, and display of the MLE system. The application included threads for (1) a command line interface to receive user inputs for transitioning between MLE illumination modes, (2) asynchronously processing data received by the frame grabber and updating the displays, and (3) communicating with the light modulation controller. Multiple buffering was employed to enable concurrent frame capturing and processing while also minimizing the risk of dropped frames. Image processing was offloaded to an Nvidia TITAN Xp graphics processing unit installed within the workstation. To maintain a well-exposed image with changes in working distance and tissue reflectance, pulse width updates were continuously sent to the light modulation controller. Updated pulse widths were computed on the host desktop using an auto-exposure technique based on an adaptation of the secant root-solving method~\cite{su2015}. A light pulse was flashed and detected in the acquired frames to synchronize the MLE system with the frame acquisition delay of the clinical system. See Supplementary Section~4 and Supplementary Fig.~5 for a detailed description of the hardware and software components and interfaces. 

The MLE system was pre-programmed with white light, topographic, laser speckle, and multispectral illumination modes. A summary of the pre-programmed illumination modes is shown in Supplementary Fig.~7.

\subsection*{Image processing}

Raw video sequences acquired by the MLE system were post-processed in Matlab~2024b (The MathWorks, Inc., Portola Valley, CA, USA). A dark frame was subtracted from the frames to remove the contribution of dark current. Deinterlacing was performed by splitting odd and even pixel rows into separate images and resizing them with bilinear interpolation. Lens distortion was removed using intrinsic parameters measured with a checkerboard target~\cite{scaramuzza2006}.

Spectral illumination frames were split into substacks each containing 8 image fields (one complete spectral cycle). Electronic and residual speckle noise was reduced by spatially averaging neighboring pixels with a 5~\texttimes{}~5 Gaussian filter ($\sigma=0.5$). The image intensity values were transformed to reflectance by correcting for the system response (diode output power, Bayer transmissivity, sensor quantum efficiency) and duty cycle (diode pulse width length). The system response for each diode was measured by imaging a reference target and computing the mean intensity of each spectral channel. To correct for movement of the tissue and colonoscope between acquisitions, each wavelength channel was registered to the 561~nm image~\cite{mattes2001}, and the affine transforms were applied to spatially align the frames.

Total oxygen saturation at each $(u,v)$ pixel was estimated from multispectral absorbance ($A$) calculated from reflectance measurements ($R$) at each wavelength ($\lambda$) using the modified Beer-Lambert law:

\begin{equation}
\begin{aligned}
A(u,v,\lambda)
&= 
-\mathrm{log_{10}}(R(u,v,\lambda)) \\
&=
\left[
c_{\scriptscriptstyle \mathrm{HbO_2}}(u,v) \cdot \epsilon_{\scriptscriptstyle \mathrm{HbO_2}}(\lambda)
+
c_{\scriptscriptstyle \mathrm{Hb}}(u,v) \cdot \epsilon_{\scriptscriptstyle \mathrm{Hb}} (\lambda)
\right] \cdot L
+ O(u,v) \\
\label{eqn:chromoFit}
\end{aligned}
\end{equation}

\noindent where $\epsilon_{\scriptscriptstyle \mathrm{HbO_2}}$ and $\epsilon_{\scriptscriptstyle \mathrm{Hb}}$ are measured molar extinction coefficients of oxy- and deoxy-hemoglobin~\cite{prahl1998}, $c_{\scriptscriptstyle \mathrm{HbO_2}}$ and $c_{\scriptscriptstyle \mathrm{Hb}}$ are the concentrations of $\mathrm{HbO_2}$ and $\mathrm{Hb}$ in the sample, $L$ is the optical pathlength, and $O$ is an additional term that accounts for attenuation by other chromophores and optical scattering~\cite{malonek1996,shonat1997}. To simplify this expression, we assumed a constant optical path length across wavelengths, allowing us to recover $c_{\scriptscriptstyle \mathrm{HbO_2}}$, $c_{\scriptscriptstyle \mathrm{Hb}}$, and $L$ from the measured reflectance spectra using linear non-negative least squares regression~\cite{clancy2015}. Total oxygen saturation was then computed as the relative concentration of oxygenation hemoglobin ($c_{\scriptscriptstyle \mathrm{HbO_2}}$) to total hemoglobin ($c_{\scriptscriptstyle \mathrm{HbO_2}}+c_{\scriptscriptstyle \mathrm{Hb}}$). Reflectance values were kept within the range (0,1) by dividing by the maximum reflectance value within the multispectral data cube plus a small constant. For the finger occlusion data (Fig.~\ref{fig:Figure2}), reflectance values were normalized with respect to the 659~nm channel~\cite{yoon2020} and converted to absorption using the Kubelka-Munk model~\cite{frank1989} with scattering modeled using parameters from~\cite{jacques2013}, and melanin was included in the unmixing using absorption values from~\cite{jacques1996}.

Color images were rendered using the spectral reflectance measurements, the transmission spectrum of a Bayer camera filter~\cite{jiang2013}, and the measured output illumination spectra from the clinical light source. Optimized weights for the rendered spectrally enhanced images were identified using gradient descent optimization, with an objective function designed to maximize the CIEDE2000 color difference~($E\delta_{00}$)~\cite{sharma2005} between 100~pixels sampled from normal and polyp tissue regions in each image. WLE and NBI images were also rendered, allowing the CIEDE2000 color difference to be computed for the same pixels as the spectral enhanced images. Prior to processing, the reflectance channels for each wavelength were rescaled to the average intensity for that wavelength across the dataset. Each pixel was normalized so that its maximum color channel value equaled 0.8 to remove the effect of brightness in the color difference calculation~\cite{waterhouse2021}. For visualization, the images were scaled so that the mean intensity of the image was equal to 0.4. See Supplementary Fig.~8) for additional implementation details.

Laser speckle-illuminated frames were post-processed to map perfusion. The MLE-HC and RGB lasers were alternately illuminated to interlace color and speckle contrast images in each full acquired frame. These frames were deinterlaced into two image stacks: laser speckle-illuminated frames originating from odd fields ($n=1,3,...,\mathrm{N}-1$) and white light-illuminated frames originating from even fields ($n=2,4,...,\mathrm{N}$). Next, the laser speckle-illuminated frames were converted to laser speckle contrast ($K$) frames using:

\begin{equation}
K = \frac{\sigma}{\langle I \rangle}.
\label{eqn:speckleContrast}
\end{equation}

Here, $\sigma$ and $\langle I \rangle$ are the standard deviation and mean within a 5~\texttimes{}~5 spatial window centered at each pixel~\cite{briers1996}. Speckle contrast frames were converted to flow contrast frames ($V$) using $V = \frac{1}{K^2}$~\cite{senarathna2019}. To improve the blood flow signal, the blood flow frames were averaged within a rolling temporal window of 15 frames. Before averaging, blood flow frames within the window were spatially aligned using registered features from the corresponding white light illuminated frames. The white light illuminated frames were registered by minimizing the mean square error between images with gradient descent optimization. 

Surface topography was recovered from directionally illuminated images using photometric stereo~\cite{woodham1980}. For simplicity, measured intensity values from the directionally illuminated frames were related by assuming a diffuse Lambertian reflectance model

\begin{equation}
I_n = \frac{\rho_d}{\pi} \cdot \hat{s}_n \cdot \hat{n},
\label{eqn:lambertian}
\end{equation}

\noindent where $I$ is the measured intensity, $\rho_d$ is a scalar term denoting the surface albedo, $\hat{s}$ is the pre-calibrated directional vector of light source $n$, and $\hat{n}$ is the surface normal to be estimated. The observed intensities and corresponding light source directions were combined with Equation~\ref{eqn:lambertian} to formulate a linear system of equations for each pixel:

\begin{equation}
\underset{\scriptscriptstyle \mathrm{N}\times 1}{I} = \underset{\scriptscriptstyle \mathrm{N}\times 3}{\hat{s}} \times \underset{\scriptscriptstyle 3\times 1}{\vec{n}}.
\end{equation}

\noindent Here, $\vec{n}$ is a non-unitary vector with magnitude $\frac{\rho_d}{\pi}$ and direction $\hat{n}$, and $\mathrm{N}$ is the number of light sources. Surface normals were estimated for every pixel within a rolling window of three image fields acquired with unique point sources. Before solving for the normals, specular reflections were inpainted, and the directional images were registered and spatially aligned by minimizing the mean square error between images with gradient descent optimization. The short working distance, wide FoV, spatially-varying anisotropic light sources located near the camera, and unknown working distance in our application introduce low-spatial frequency error to the recovered surface normals. This error was removed using a high-pass filter ($\sigma=150$), which was applied by convolving the surface normal map with a low spatial frequency filter and subtracting the result from the unfiltered surface normal map~\cite{parot2013}. To reconstruct a height map, the filtered surface normals were projected to gradient space and integrated using a multigrid solver for the Poisson equation~\cite{farneback2007,simchony1990}.

\subsection*{Spectral fidelity evaluation}

The spectral fidelity of MLE was characterized by imaging a Macbeth Color Chart (ColorChecker Classic Nano; X-Rite, Inc., US). Reference measurements of a diffuse polytetrafluoroethylene (PTFE) target (PMR10P1; Thorlabs, Inc., Newton, NJ, US) were acquired and corrected using ground truth reflectance spectra provided by the manufacturer. The PTFE target was then replaced with the color target and sample measurements ($I$) were acquired. The sample measurements were normalized by dividing them by the corrected white reference images to compensate for spatial illumination nonuniformity and system response, including variations in diode output power, Bayer filter transmissivity, and sensor quantum efficiency. 60~\texttimes{}~60~pixel regions of interest were cropped from each patch on the color chart and used to calculate the mean reflectance value at each wavelength. Ground truth reference spectra for each patch were acquired with a CCD spectrometer (CCS200; Thorlabs, Inc., Newton, NJ, US), a fiber-coupled stabilized broadband light source (SLS201L; Thorlabs, Inc., Newton, NJ, US), and a bifurcated reflection probe (RP20; Thorlabs, Inc., Newton, NJ, US).

\subsection*{Finger occlusion experiment}

A blood pressure cuff (Manual Inflate Blood Pressure Kit; Walgreens, Deerfield, IL, US) was applied to the forearm of a healthy volunteer. The volunteer's hand was depressed into a soft clay (0716623001735; Crayola, Easton, PA, US) disk and the endoscope tip was mounted above the index finger to minimize motion. A 420~sec video was recorded with multispectral illumination. After imaging 120 seconds of baseline video, the pressure cuff was inflated to 300~mmHg to occlude blood flow for 180~sec. The pressure cuff was then released and an additional 120~sec of video was recorded. The protocol for this study was approved by the Johns Hopkins Institutional Review Board (IRB00344195). All participants were over the age of 18 years and provided written informed consent prior to participation. This research was performed in accordance with the Declaration of Helsinki.

\subsection*{Microfluidic flow phantom experiment}

A microfluidic flow phantom was fabricated to assess the sensitivity of MLE to differences in flow velocity. The phantom included 500~\textmu{}m wide and 150~\textmu{}m tall channels manufactured from polydimethylsiloxane (PDMS) doped with titanium dioxide (TiO\textsubscript{2}) and India ink for optical scattering ($\mu_{s}'~\sim~2.0~mm^{-1}$, $\lambda$~=~650~nm) and absorption ($\mu_{a}~\sim~0.01~mm^{-1}$, $\lambda$~=~650~nm), respectively. The flow channels were sealed with a 100~\textmu{}m thick PDMS membrane also doped with TiO\textsubscript{2} and India ink using the process described in~\cite{mckay2020}. A syringe pump (98-5457; Harvard Apparatus, Cambridge, MA, US) was used to pump the polystyrene microsphere solution through the phantom to simulate blood flow. A 1.0~mL solution of deionized water and 0.943~\textmu{}m diameter polystyrene microspheres (PS03N; Bangs Laboratories Inc, Fishers, IN, US) was mixed 13.86:1 to achieve a reduced scattering coefficient comparable to that of human blood ($\mu_{s}'~\sim~2.0~mm^{-1}$, $\lambda$~=~633~nm)~\cite{friebel2006}. The suspension was sonicated and transferred to a 1~mL syringe. The solution was continually mixed using a stir rod within the syringe and a rotating magnet to maintain a homogeneous suspension during the experiment. The syringe was connected to the input port of the microfluidic phantom using PEEK tubing (1569L; IDEX Corporation, Lake Forest, IL, US).

The microsphere solution was pumped through the channel at velocities ranging from 0.4~mm/s to 2.2~mm/s and exposure times ranging from 3~ms to 13~ms. One hundred frames were acquired per condition, with a 120 s pause between acquisitions to ensure flow stabilization. To statistically assess whether the relationship between contrast and velocity differed across exposure times, we fitted an ANCOVA model with an interaction term (contrast~$\sim$~velocity~\texttimes~exposure~time). The significance of the interaction term (velocity~\texttimes~exposure~time) was evaluated via an ANOVA F-test, with $\alpha$~=~0.05.

\subsection*{Silicone adenoma phantom fabrication and imaging}

The surface measurement capability of MLE was evaluated using a phantom colon model with a known three-dimensional shape~\cite{bobrow2023}. A mold for the model was 3D-printed using a Form 3 printer (FormLabs, Somerville, MA, US) at a resolution of 25~\textmu{}m. The mold was cast with silicone (Dragon Skin\texttrademark{}, Smooth-On, Inc., Macungie, PA, US), with pigments (Silc Pig\texttrademark{}, Smooth-On, Inc., Macungie, PA, US) added to replicate the color of gastrointestinal tissue. Images of the phantom model were manually registered to the ground truth model via a custom graphical user interface. Ground truth depth frames were rendered for comparison with those measured by MLE.

\subsection*{Soft palette and ventral tongue imaging}

To test the MLE system in an imaging environment with similar properties to that of colon tissue, the oral mucosa of the ventral tongue and the soft palate were imaged in three human participants using a protocol approved by the Johns Hopkins Institutional Review Board (IRB00279890). Subjects were handed the colonoscope and instructed to slowly insert the scope tip into the oral cavity. Approximately 10~s of data was recorded per video. The participant was then given a moment to rest, and the process was repeated four additional times. All participants were over the age of 18 years and provided written informed consent prior to participation. This research was performed in accordance with the Declaration of Helsinki.

\subsection*{Clinical study design}

A first-in-human pilot clinical study was conducted to assess the ability of MLE to enhance lesion contrast in colonoscopy. The study enrolled patients who were scheduled for a screening or surveillance colonoscopy. All participants were over the age of 18 years and were capable of giving informed consent. Patients with bleeding or hemostasis disorders, patients taking anticoagulants such as Warfarin and Clopidogrel, and patients with colitis or active bleeding were excluded from the study.

After obtaining informed consent, the patient was escorted to the procedure room and placed under conscious sedation. The MLE system was then rolled into position next to the endoscopy tower, powered on, and the optical fibers were connected to the MLE fiber ports on the retrofitted scope. The shutters on the laser source were closed, and the lasers were toggled on and allowed to warm up. The retrofitted colonoscope was inserted through the anus and advanced to the cecum. As the colonoscope was withdrawn, the endoscopist examined the colon for polyps. Upon discovery of a polyp, the endoscopist alerted the study team. To avoid extending the procedure length, polyps were imaged with MLE research modalities while the resection device and associated materials were gathered and inserted through the instrument channel to the tip of the scope (typically 30~s).

Upon discovery of a polyp, the CV-190L light source was toggled off. After sending a synchronization pulse, the MLE light source was toggled on, and data was recorded using each of the illumination modes. Once resected, the polyp was assigned a unique sample identifier, and ground truth histology and location was retrieved from the subject's medical record post-procedure. Frames for each illumination mode were isolated and cropped so that the images from exhibited a similar FoV. The polyps were manually segmented in the extracted frames by an experienced endoscopist.

Participants were not prospectively assigned to an intervention intended to evaluate effects on health outcomes. During routine colonoscopy, polyps were first identified using standard-of-care clinical methods. After identification, images were acquired using an investigational imaging system for research purposes only. The investigational images were not used for clinical decision-making and did not influence patient management, diagnosis, or treatment. The study therefore represents an observational imaging study in which research images were collected alongside standard clinical care and analyzed retrospectively. This research was performed in accordance with the Declaration of Helsinki. The study was approved as a Non-Significant Risk protocol by the Johns Hopkins Institutional Review Board (IRB00279918).

\subsection*{Clinical statistical analysis}
To assess functional differences ($\mathrm{StO_2}$ and flow), the mean values of normal and lesion tissue regions were computed for each sample. CIEDE color differences were analyzed by comparing Sim-WLE to Sim-NBI and Sim-SE. For RMS contrast, WLE-Hue and WLE-Saturation were compared against SE-Hue, SE-Saturation, height, flow, and $\mathrm{StO_2}$. For all analyses, only mutually available paired samples were included; samples not present in both groups were excluded. Normality of the paired differences was evaluated using the Shapiro–Wilk test. When normality was satisfied, paired two-sided \textit{t}-tests were performed. For non-normal comparisons, significance was assessed using the Wilcoxon signed-rank test, and 95\% confidence intervals for the mean paired difference were estimated using a percentile bootstrap (10,000 resamples). For RMS contrast analyses, Holm–Bonferroni corrections were applied separately to the WLE-Hue and WLE-Saturation comparison families to control the family-wise error rate.

\section*{Data availability}

The data acquired and analyzed in this work is available online at: \url{https://doi.org/10.7281/T1TQIZJ9}.

\section*{Code availability}

The code used in this study for image acquisition and processing is available online at: \url{https://doi.org/10.7281/T1TQIZJ9}.

\section*{Acknowledgments}

This study was funded by the National Institutes of Health Trailblazer Award (R21 EB024700), the National Science Foundation (NSF) Graduate Research Fellowship Program (DGE-1746891), and the NSF CAREER Award (2146333). The funder played no role in study design, data collection, analysis and interpretation of data, or the writing of this manuscript. The authors thank Terry Mistalski (HCS Solutions Pro), Christopher Perry (Cantel Medical Corporation), Tom Arcand (Cantel Medical Corporation), Daniel Pferdeort (Cantel Medical Corporation), Kiona Williams (Johns Hopkins Hospital Central Sterile Processing), and Elaine Nowakowski (Johns Hopkins Hospital Epidemiology and Infection Control) for assistance with infection prevention measures. The authors also thank Greg McKay for sharing photographs of the system included in Fig. 1.

\section*{Author contributions}

Optical design, device fabrication, acquisition software, and data processing: TLB and NJD; clinical study design: TLB, SN, and NJD; data collection: TLB, MG, AAS, and SN; data curation: TLB, SA, and SN; manuscript preparation: TLB, SA, and NJD.

\section*{Competing interests}

NJD and TLB are co-inventors of two issued patents (US11971960B2; US12361097B2) relating to the technology presented here. SN is a consultant for Boston Scientific Corp, Olympus, Fujifilm, and Neptune Medical. The remaining authors declare no competing interests.



\clearpage

\clearpage

\clearpage

\clearpage

\clearpage

\clearpage

\makeatletter
\renewcommand{\figurename}{Supplementary Figure}
\renewcommand{\tablename}{Supplementary Table}

\begin{center}
\textbf{\Large Supplementary Material}
\end{center}

\section{Retrofitted clinical colonoscope fabrication} \label{sec:sup_scope}

\begin{figure}
\centering
\includegraphics{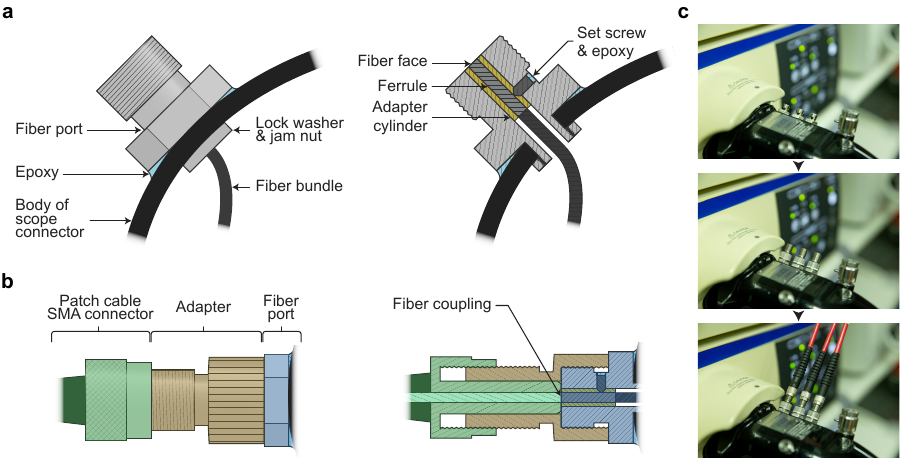}
\caption{\textbf{Scope connector modifications.} \textbf{a} Three externally threaded SMA fiber ports were installed through the connector end of the colonoscope. The MLE fiber bundle ferrules were passed from the scope’s internal lumen through the fiber ports and secured. External surfaces were smoothed with epoxy to minimize crevices that could harbor bacteria. \textbf{b} Custom internal-to-external SMA-threaded adapters were machined for coupling SMA-terminated fiber patch cables to the MLE fiber bundles. \textbf{c} Images of the MLE fiber ports, SMA-threaded adapters, and fiber patch cables.}
\label{fig:FigureS1}
\end{figure}

The internal fiber optic light guide of an adult video colonoscope was replaced with a custom fiber optic bundle. This custom bundle maps half of the fibers from the three distal sources to the clinical white light source, and the other half are mapped to three new 1~mm diameter MLE fiber bundles at the connector end of the scope. We found that the white light source was sufficiently bright to maintain standard-of-care illumination with the outer half of its illumination fibers removed (original cross-sectional area~=~4.6~mm\textsuperscript{2}, new cross-sectional area~=~2.3~mm\textsuperscript{2}). To make the internal MLE fiber bundles accessible for optical coupling, the fiber bundle ferrules were mounted within stainless steel fiber ports added to the colonoscope (Supplementary Fig.~\ref{fig:FigureS1}a). Three 7.4~mm diameter holes, spaced 13~mm center-to-center, were drilled through the scope connector body for the fiber ports. SMA fiber ports (ADASMA; Thorlabs, Inc., Newton, NJ, US) were installed through each hole and fastened internally using lock washers, jam nuts, and high-temperature epoxy (Cyberpoxy 5895; H.B. Fuller, Saint Paul, MN, US). Each fiber bundle was secured into a brass cylindrical adapter with an outer diameter equal to the inner diameter of the fiber port. A small set screw secured each fiber adapter within the fiber port. Crevices around each fiber port were filled with epoxy to minimize surfaces that could harbor bacteria or fluid during scope cleaning. Multimode fiber patch cables (MMF1) were used to transmit light from the illumination source, and these cables were optically butt-coupled with the MLE fiber bundles by applying light pressure between the fiber faces with custom-machined stainless steel SMA adapters (Supplementary Fig.~\ref{fig:FigureS1}b).

To reduce the risk of infection arising from the use of a retrofitted colonoscope, several precautionary measures were taken. The manufacturer of the automated endoscope reprocessors (AER) used by Johns Hopkins (Cantel Medical Corporation, NJ, USA) conducted a design review and a chemical analysis of the materials used to retrofit the colonoscope. The manufacturer approved the retrofitted scope for reprocessing in the following AER configurations using the standard instructions for use (IFU):

\begin{itemize}
\item ADVANTAGE PLUS and ADVANTAGE PLUS Pass-Thru AERs w/ a 2-8-611 hookup and the 1-xx-603 parameter set
\item DSD EDGE and DSD-201 AERs w/ DSD-110-HU0190 and DSD-110-HU0102 hookups
\item CER OPTIMA AERs w/ CLM-110-HU0108, CLM-110-HU0124, CLM-110-HU0129, and CLM-110-HU0130 hookups
\end{itemize}

All patients included in the study were longitudinally monitored post-procedure by the hospital infection control unit, and no cases of post-endoscopic infection or other adverse events were reported.

\section{Multi-contrast laser illumination source optical design} \label{sec:sup_lightsource}

The optical layout and component list for the multi-contrast illumination source are reported in Supplementary Fig.~\ref{fig:FigureS2} and Supplementary Table~\ref{tab:TableS1}, respectively.

The Red-Green-Blue (RGB) laser units were constructed from 450~nm (LD-450), 525~nm (LD-525), and 637~nm (LD-637) multimode laser diodes. The laser diodes were collimated using aspheric lenses (L1), and the slow axes were magnified using cylindrical lens pairs (CLP) to attain symmetrical beam shapes. Silver (M1) and dichroic (DM1, DM2) mirrors were employed to consolidate the beams into a coaxial path. A wave rotator plate (HWP1) was used to rotate the 637~nm diode's polarization angle by 90\textdegree~so that all three diodes exhibited S-polarization.

The light source also includes 405~nm (LD-405), 462~nm (LD-462), 543~nm (DPSS-543), 561~nm (DPSS-561), and 659~nm (LD-659) laser sources. Similar to the RGB units, aspheric lenses (L3, L5) and cylindrical lens pairs (CLP) were used to collimate the multimode laser diodes (LD-405, LD-462, LD-659). The 543~nm and 561~nm sources have coherence lengths long enough to generate laser speckle noise. To reduce this noise, both sources were coupled into multimode fibers (MF2), passed through a fiber oscillator (FO), and re-collimated using fixed fiber collimators (L4). Dichroic mirrors (DM3-DM6) with sharp cut-on wavelengths and a silver mirror (M2) combined the five lasers into a coaxial beam path. To improve illumination homogeneity and reduce shadows, the beam was split and coupled with RGB units 1 and 2 so that multispectral illumination was emitted from $\vec{L}_1$ and $\vec{L}_2$ simultaneously. This splitting and coupling was achieved with a 50:50 beam splitter (BS) and polarized beam splitters (PBS). All of the lasers exhibited P-polarization, causing them to be primarily reflected by the polarized beam splitters for coupling into the RGB beam paths. Any residual light not reflected by the polarized beam splitters was captured by beam blocks (BB).

A 639~nm single longitudinal mode laser (coherence length~>~40~m) was combined with RGB module 3 to generate laser speckle for flow contrast imaging. To modulate the output power of the laser, the beam's polarization angle was rotated to vary the ratio of transmitted and reflected light by the polarization beam splitter. A half-wave plate (HWP2) was mounted to a direct drive rotation mount (RM) to control the beam's polarization angle programmatically.

The beams were steered using pairs of broadband dielectric mirrors (M2) into aspheric condenser lenses (L2). These lenses focused and coupled the light into 1~mm diameter multimode fiber optic patch cables (MF1). To ensure complete filling of the fibers and to minimize the risk of fiber damage, the fiber faces were positioned at approximately 1~mm before the back focal length of the aspheres. Custom-machined stainless steel SMA adapters were used to screw together the fiber patch cables and the MLE fiber ports, applying light pressure between the fiber faces for optical butt-coupling Supplementary Fig.~\ref{fig:FigureS1}b). Mechanical shutters (SHTR) were placed before the aspheres so the system could be warmed up without emitting light from the MLE illumination source.

The laser diodes were powered by drivers equipped with transistor–transistor logic (TTL) switching for pulse width modulation (PWM) exposure control. The TO-can diodes (LD-\textasteriskcentered) were powered by single channel constant current drivers (LPLDD-5A-24V-TP-H; Opt Lasers, Warszawa, PL) supplied by a 24~V power supply (HRP-300-24; MEANWELL Enterprises Co., Ltd., New Taipei City, TW). Laser diode temperature was regulated by PID-controlled thermoelectric cooler (TEC) drivers (TEC-20A-25V-PID-H; Opt Lasers, Warszawa, PL) powered by a 7.5~V power supply (HRP-100-7.5; MEANWELL Enterprises Co., Ltd., New Taipei City, TW). The DPSS lasers were powered by manufacturer-calibrated drivers/power supplies (PSU-H-LED/PSU-H-FDA; CNI Laser, Changchun, CN) that include dials for tuning the output power. A summary of the laser diode packages and drivers is reported in Supplementary Table~\ref{tab:TableS2}.

The drive current for each laser diode was tuned so that the total output power emitted by the scope with all diode channels enabled was less than the maximum output power emitted by an unmodified colonoscope and clinical light source. The maximum clinical illumination power was measured by setting the clinical light source to manual illumination mode and the light output to the highest output setting. A broadband optical power meter (S470C; Thorlabs, Inc., Newton, NJ, US) was placed at the colonoscope's distal tip, and the optical output power was measured as 330~mW. Using the optical power meter, the output power of each laser diode was tuned to the values listed in Supplementary Table~\ref{tab:TableS2}, which combined are less than the maximum clinical output power.

To correct for drift and instability in the laser output power, coverglass slips (CG) were used to pick off a small percentage ($\sim$3\%) of light from each beam path for continuous monitoring. The light was measured by a photodiode (PD) mounted to a printed circuit board (PCB) fabricated by JiaLiChuang~PCB (JLCPCB; Shenzhen, CN). Neutral density filters (ND) reduced the power to fall within the dynamic range of the photodiodes, and ground glass diffusers (DIFF) spread the light to reduce sensitivity to alignment. The photodiode PCB, diffuser, neutral density filter, and cover glass were combined with a 3D-printed mount and a cage adapter (CP30Q; Thorlabs, Inc., Newton, NJ, US). The electrical schematic and images of the output power monitoring unit are depicted in Supplementary Fig.~\ref{fig:FigureS3}.

All optical components were mounted to a 24~\texttimes{}~36~in optical breadboard (B2436F; Thorlabs, Inc., Newton, NJ, US) installed on a rolling cart (POC001; Thorlabs, Inc., Newton, NJ, US) for transport to and from the clinic. An enclosure was constructed from extruded aluminum rails and laser-cut acrylic panels to contain stray light and to protect the equipment from fluid splashes and spills. Fans were installed in the walls of the enclosure to extract heat from the enclosure. All electronics were powered by a medical-grade isolation transformer (AN120411; Toroid, Salisbury, MD, US) installed in the cart's base.

\section{Light modulation controller} \label{sec:sup_lmc}

Pulse width modulation (PWM) of the laser diodes was managed by a custom light modulation controller PCB. A circuit schematic and photograph of the PCB are shown in Supplementary Fig.~\ref{fig:FigureS4}. The PCB connects to the analog separate video (S-video) output of the Olympus CV-190 video processor via a 4-pin mini-DIN cable. A low-pass filter with a cut-off frequency of 500~kHz was included to remove noise from the raw video signal. The PCB included a chip (LM1881; Texas Instruments, Dallas, TX, US) that extracts the odd/even field vertical synchronization pulses from the filtered S-video input signal. This signal was tied to a digital interrupt pin on a microcontroller unit (MCU; Teensy 4.0; PJRC, Sherwood, OR, US) to synchronize the triggering of illumination pulses with the start of each frame acquisition. Digital output pins on the MCU modulated the laser diodes. Logic level shifters (74HCT245N; Texas Instruments, Dallas, TX, US) converted the digital output pin signals from 3.3~V to 5.0~V for compatibility with the laser diode drivers. Quad timers within the MCU were used to time the pulse width lengths of the diode channels to control the exposure times. An additional quad timer was used to time the reading of the analog input voltages from the output power monitoring units. The measurements were acquired 50~\textmu{}s after the rising edge of the PWM signals --- a value greater than the 35~\textmu{}s rise time of the photodiodes within the laser power monitoring units.

\section{System architecture} \label{sec:sup_arch}

A wireframe diagram outlining the hardware interfaces between the clinical and MLE systems is shown in Supplementary Fig.~\ref{fig:FigureS5}.

\textbf{Clinical hardware:} The clinical endoscopic imaging system consists of a video processor (CV-190; Olympus America Inc., Center Valley, PA, US), xenon light source (CLV-190; Olympus America Inc., Center Valley, PA, US), and two display monitors. The connector end of the colonoscope plugs into the clinical light source and video processor for optical coupling with the arc lamp and transmission of the CCD signal to the video processor. The colonoscope acquires images at a frequency and resolution of 29.97 interlaced frames per second and 1080~\texttimes{}~1350~pixels, respectively. Odd and even rows of pixels, also called image fields, are acquired by the colonoscope in an alternating fashion at a frequency twice the frame rate (59.94 fields per second). Acquired frames are transmitted to the video processor for post-processing and output via the display ports. An serial digital interface (SDI) video output from the video processor is connected to Display Monitor 1 for visualization by the clinician. The clinical system includes a second display monitor (Display Monitor 2) that is not typically utilized during screening colonoscopy procedures. This monitor was repurposed to display a blended video feed from the MLE system for easier visualization while imaging in MLE illumination modes. To reduce post-processing of the image data, the Electronic Shutter (ES) and Auto Gain Control (AGC) settings were disabled, and the Edge Enhancement mode was set to \textit{A0} in the user settings of the CV-190 video processor.

\textbf{MLE hardware:} Control of the MLE system was managed by a Dell Precision 5820T Workstation (Intel i7-9800X 3.80GHz 8-core Central Processing Unit (CPU); 64.0~Gb physical memory) running Microsoft Windows~10. An Orion HD frame grabber (Matrox Imaging, Montreal, CA) captured and stored uncompressed video from the video processor. Image processing was offloaded to an Nvidia TITAN Xp graphics processing unit (GPU) installed within the workstation. The system also included a portable monitor for displaying a post-processed computational video feed that was visualized by the research team. Communication with the light modulation controller and direct drive rotation mount occurred over native universal serial bus (USB).

\textbf{Data acquisition software:} A custom C++ application was developed to enable real-time data acquisition and control of the MLE system. Independent application threads were launched for (1) a command line interface (CLI) to receive user inputs for transitioning between MLE illumination modes, (2) asynchronously processing data received by the frame grabber and updating the displays, and (3) communicating with the light modulation controller.

Multiple buffering was employed to enable concurrent frame capturing and processing. Frames were continuously acquired by the frame grabber and transmitted to the host desktop, where they were sequentially stored in a list of image buffers pre-allocated in physical memory. In parallel, the frame processing thread sequentially copied frames from the image buffer list and processed them, freeing spots in it. By continuously grabbing and storing in a list of image buffers, the system was robust to changes in operating system overhead, preventing frames from being dropped as long as the list was never filled. Once copied by the frame processing thread, the frame was written to the workstation's solid-state drive in an uncompressed format. The frame was then copied to GPU memory for image processing.

Image processing was executed on the GPU for accelerated computation. First, each frame was deinterlaced by separating the odd and even fields into separate arrays. The fields were then resized to the original image height using linear interpolation. The deinterlaced frames were processed and rendered to Display Monitor 1 and Display Monitor 2. Next, the mean intensity of each color channel for each deinterlaced frame was computed using a highly parallelized sum reduction algorithm implemented using the Compute Unified Device Architecture (CUDA) platform. If a cap attachment was used with the colonoscope, a circular mask was applied to exclude pixels occluded by the cap from the computed image statistics.

To maintain a well-exposed image with changes in working distance and tissue reflectance, the computed mean intensity values were used to generate updated pulse width lengths for every diode. Pulse width updates were computed using an auto-exposure technique based on an adaptation of the secant root-solving method. The algorithm assumes that image brightness and exposure time (pulse width length) are related by an unknown, monotonically increasing function. Convergence towards the function's root (the target mean intensity) is achieved by iteratively approximating the function with a secant line and solving for the line's root. This technique has several advantages, including fast convergence when applied to nonlinear functions (such as gamma-corrected images) and the ability to recover from erroneous values (caused by light modulation controller malfunctions, for example).

Given a mean intensity value $I_{\scriptscriptstyle n}$ illuminated with pulse width length $P_{\scriptscriptstyle n}$, an updated pulse width length $P_{\scriptscriptstyle n+1}$ can be computed using

\begin{equation} 
	P_{\scriptscriptstyle n+1} = \frac{[I_{\mathrm{\scriptscriptstyle max}}-I_{\scriptscriptstyle n}] \cdot P_{\scriptscriptstyle n} \cdot P_{\mathrm{\scriptscriptstyle max}}} {[I_{\mathrm{\scriptscriptstyle target}}-I_{\scriptscriptstyle n}] \cdot P_{\scriptscriptstyle n} + [I_{\mathrm{\scriptscriptstyle max}}-I_{\mathrm{\scriptscriptstyle target}}] \cdot P_{\mathrm{\scriptscriptstyle max}}},
	\label{eqn:mle_auto-exposure}
\end{equation}

\noindent where $I_{\mathrm{\scriptscriptstyle max}}$ is the upper bound of the mean intensity (255), $P_{\mathrm{\scriptscriptstyle max}}$ is the maximum pulse width length (14~ms), and $I_{\mathrm{\scriptscriptstyle target}}$ is the target mean image intensity (128). Pulse widths were restricted to a maximum length of 14~ms to avoid bleeding into the next image field acquired by the colonoscope. $I_{\scriptscriptstyle n}$ was set to the mean intensity value from the color channel with the highest Bayer transmissivity for the laser diode being updated. For white light illumination, the mean intensity values for all three color channels were averaged to generate a single value. While imaging with light from the high coherence laser, pulse width lengths were linearly mapped to angles formed by the fast axis of the half wave plate and the vertical axis of the laser (0~ms~\textrightarrow{}~0\textdegree, 14~ms~\textrightarrow{}~45\textdegree). Updated pulse width lengths and half-wave plate rotation angles were continuously written to a log file to record all illumination powers used for each frame.

Updated pulse width lengths for the odd and even fields were sent to the light modulation controller in 64~byte packets containing a unique frame ID (4~bytes), 16~bit pulse width lengths for each diode for the odd field (30~bytes), and 16~bit pulse width lengths for each diode for the even field (30~bytes). Measured laser output powers were sent from the light modulation controller to the host in 16~byte packets containing the received frame ID (4~bytes), 16~bit power measurements for the odd field (6~bytes), 16~bit power measurements for the even field (6~bytes). First-in-first-out (FIFO) buffers were pre-allocated on the host and the MCU for sending/receiving packets asynchronously to prevent blocking. Upon connection with the host desktop, the light modulation controller pulse width buffer was populated with a set of empty packets equal to the length of the image buffer list on the host desktop, effectively causing it to work several frames behind the host desktop. This ensured that the light modulation controller would not run out of pulse width lengths even when the host desktop fell behind on frame processing. An image buffer list 10 frames in length was empirically determined to provide an optimal balance between preventing dropped frames and minimizing latency in auto-exposure response.

Upon sending a pulse width packet to the light modulation controller, several frame acquisitions occur before the frame illuminated by the pulse width packet is seen and processed by the frame processing thread. This delay depends on the length of the pulse width buffer, the delay between frame acquisition and output by the video processor, and the delay in frame acquisition by the video capture card. Equation~\ref{eqn:mle_auto-exposure} accepts pairs of pulse width lengths and corresponding image mean intensities, so the delay between sending a pulse width and processing the corresponding image on the host desktop must be determined. To measure the delay, a single illumination synchronization pulse was sent to the light modulation controller. A counter was incremented with each frame processed by the frame processing thread, terminating once the frame containing the synchronization pulse was detected. The delay length varied between clinical systems, so a new measurement was taken during each imaging session. The delay was typically 23-24 frames, resulting in an auto-exposure response delay equal to approximately 0.8~s.

\section{Supplementary movies}

\textbf{Supplementary Movie 1:} Total oxygen saturation measurements of an index finger during an occlusion trial captured using MLE.

\textbf{Supplementary Movie 2:} Laser speckle contrast imaging of vasculature in the soft palate captured using MLE in freehand operation.

\textbf{Supplementary Movie 3:} Topographic imaging of the ventral tongue captured using MLE in freehand operation.

\textbf{Supplementary Movie 4:} 4~mm tubular adenoma in the ascending colon imaged \textit{in~vivo} with WLE and MLE’s topographic, blood flow, oxygenation, and spectral enhancement (with unsharp masking) modes.

\textbf{Supplementary Movie 5:} 5~mm tubular adenoma in the descending colon imaged \textit{in~vivo} with MLE’s topographic, blood flow, oxygenation, and spectral enhancement (with unsharp masking) modes.

\textbf{Supplementary Movie 6:} 3~mm tubular adenoma in the cecum imaged \textit{in~vivo} with MLE’s topographic enhancement mode.

\clearpage

\clearpage
\begin{figure}
\centering
\includegraphics{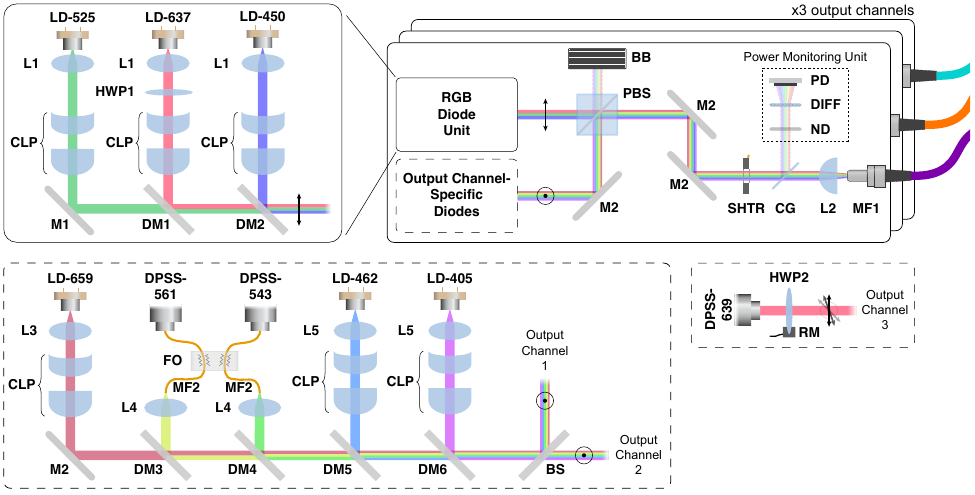}
\caption{\textbf{Laser illumination source optical schematic.} Three RGB laser units are used for white light and directional illumination modes. An additional 5 wavelengths are coupled into output channels 1 and 2 to expand the diversity of wavelengths available for multispectral imaging. A high-coherence laser is coupled into output channel 3 for speckle illumination. BB: Beam Block, BS: Beam Splitter, CG: Glass Coverslip, CLP: Cylindrical Lens Pair, DIFF: Diffuser, DM: Dichroic Mirror, DPSS: Diode-Pumped Solid-State Laser, FO: Fiber Oscillator, HWP: Half-Wave Plate, L: Aspheric Lens, LD: Laser Diode, M: Mirror, MF: Multimode Fiber, ND: Neutral Density Filter, PBS: Polarized Beam Splitter, PD: Photodiode, RM: Rotation Mount, SHTR: Shutter. A detailed component list is provided in Supplementary Table~\ref{tab:TableS1}.}
\label{fig:FigureS2}
\end{figure}

\clearpage
\begin{figure}
\centering
\includegraphics{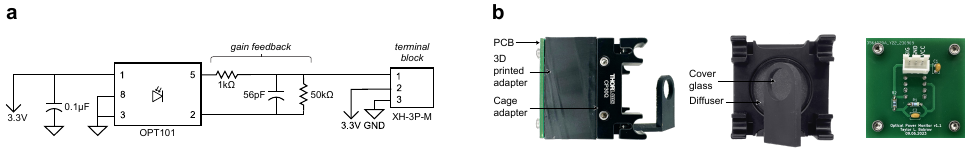}
\caption{\textbf{Laser diode output power monitoring unit.} (a) Electrical schematic. (b) Annotated photographs of the unit assembly. A cover glass slide picks off a small fraction of light for measurement by the photodiode. A ground glass diffuser is mounted between the coverglass and the photodiode to reduce the unit's sensitivity to misalignment.}
\label{fig:FigureS3}
\end{figure}

\clearpage
\begin{figure}
\centering
\includegraphics{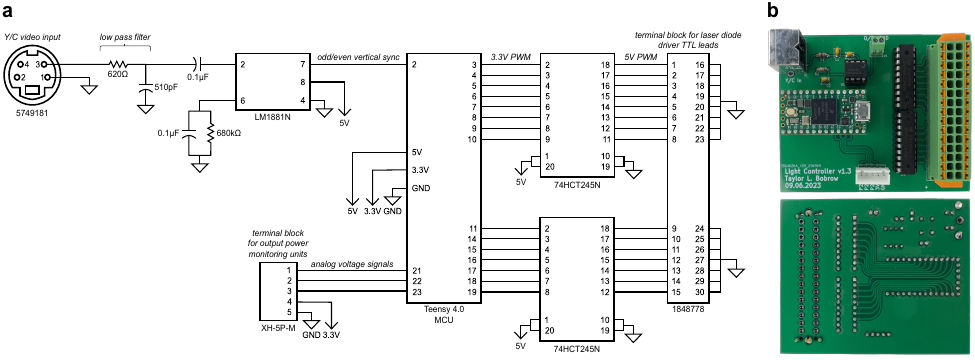}
\caption{\textbf{Light modulation controller.} \textbf{a} Electrical schematic. \textbf{b} Front (top) and back (bottom) photographs of the custom-fabricated PCB.}
\label{fig:FigureS4}
\end{figure}

\clearpage
\begin{figure}
\centering
\includegraphics{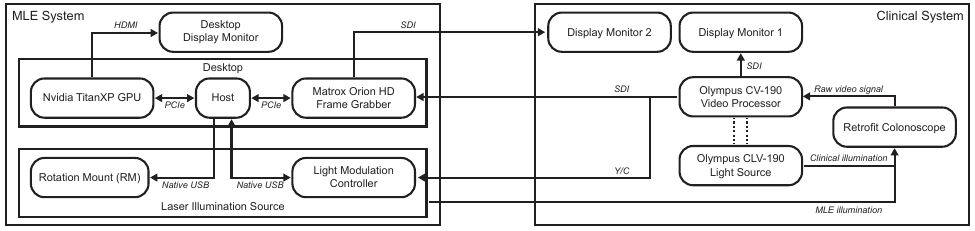}
\caption{\textbf{Wireframe diagram of MLE and clinical system architecture.}}
\label{fig:FigureS5}
\end{figure}

\clearpage
\begin{figure}
\centering
\includegraphics{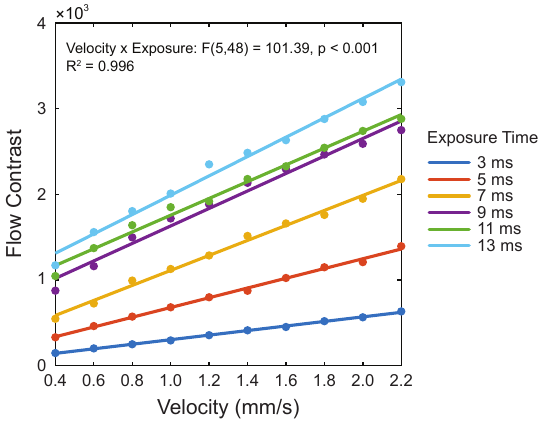}
\caption{{\textbf{Flow contrast increases with velocity at a rate dependent on exposure time}. Scatter points show mean contrast at each velocity for each exposure time. Lines show fitted values from an ANCOVA model (contrast~$\sim$~velocity~\texttimes~exposure time). The interaction between velocity and exposure time was significant (F(5, 48)~=~101.39, $p$~<~0.001, $R^2$~=~0.996), indicating that longer exposure times amplify the rate at which contrast increases with velocity.}}
\label{fig:FigureS6}
\end{figure}

\clearpage
\begin{figure}
\centering
\includegraphics{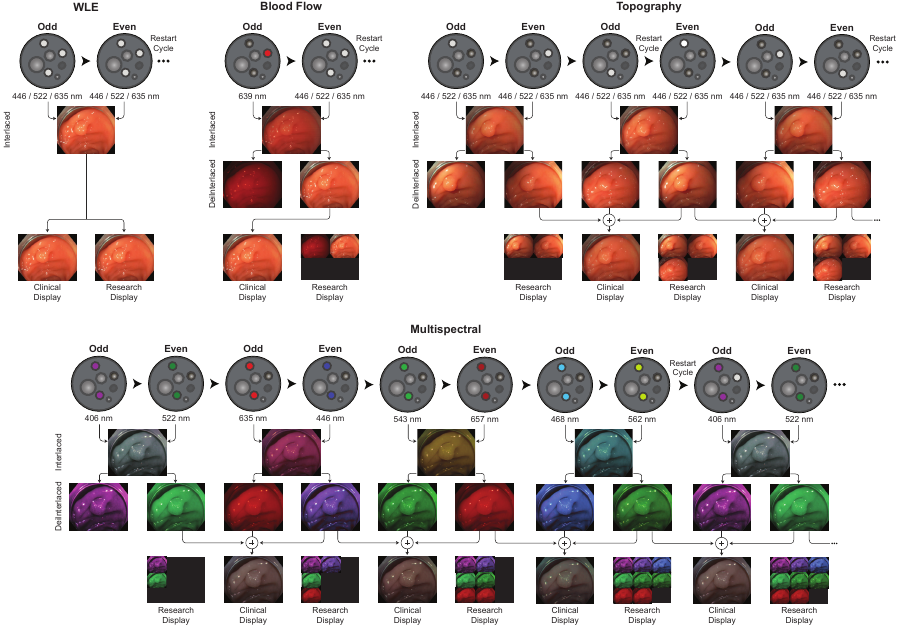}
\caption{\textbf{MLE illumination modes.} The MLE system was pre-programmed to cycle between white light, topographic, laser speckle, and multispectral illumination modes. Each mode was defined with a laser diode sequence, relative diode intensities, color channels used for auto-exposure, and post-processing steps for research and clinical video display outputs.}
\label{fig:FigureS7}
\end{figure}

\clearpage
\begin{figure}
\centering
\includegraphics{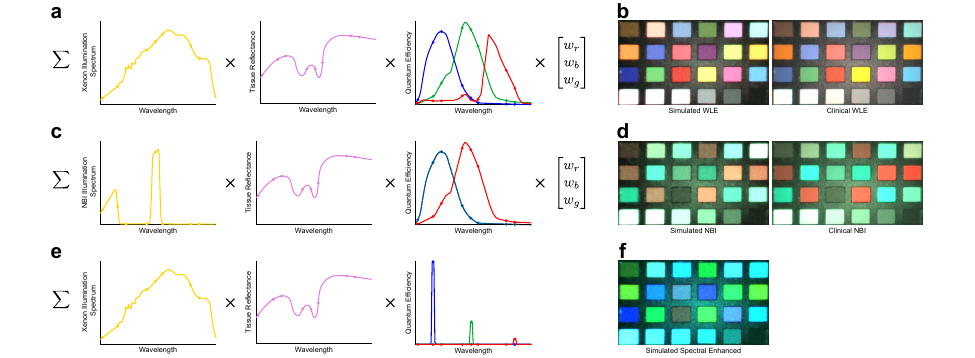}
\caption{\textbf{Color image simulation.} \textbf{a} White light endoscopy (WLE) color simulation process. The xenon illumination spectrum from the Olympus CLV-190 light source is attenuated by reflectance spectra acquired with MLE and the transmission spectrum of a commercial Bayer camera filter. A correction factor ($\omega$) is applied to each color channel to compensate for differences in the total energy between the complete and sparsely sampled illumination/camera transmissivity. The correction factor is computed by dividing the sum of the product of the full illumination spectra and color channel by the eight sparsely sampled spectra. \textbf{b} Ground truth and simulated WLE images. The simulated image was linearly color balanced using the ground truth image as a reference. \textbf{c} Narrow band imaging (NBI) color simulation process. The NBI illumination spectrum from the Olympus CLV-190 light source is attenuated by reflectance spectra acquired with MLE and the NBI color channel weightings. Intensity values from the blue Bayer channel are assigned to the green/blue image channels, and intensity values from the green Bayer channel are assigned to the red image channel. \textbf{d} Ground truth and simulated NBI images. The simulated image was linearly color balanced using the ground truth image as a reference. \textbf{e} Spectral enhanced color simulation process. The xenon illumination spectrum is attenuated by reflectance spectra acquired with MLE and camera transmission spectra obtained from the optimizer. \textbf{f} Simulated spectral enhanced image.}
\label{fig:FigureS8}
\end{figure}

\clearpage
\begin{table}
\caption{Laser illumination source optical components}
\centering
\begin{tabular}{p{1.55cm} p{12.9cm}}
\toprule
\textbf{Key}    & \textbf{Component}                                                    							\\
\midrule
BB              & Beam Block (LB1; Thorlabs, Inc., Newton, NJ, US)                                            		\\
BS              & 50:50 UVFS Plate Beam Splitter (BSW26; Thorlabs, Inc., Newton, NJ, US)                      		\\
CG              & Glass Coverslip, $\diameter$~=~12~mm (26023; Ted Pella Inc., Redding, CA, US)                    	\\
CLP             & Cylindrical Lens Pair, M~=~3\texttimes{} (001024; Opt Lasers, Warszawa, PL)           			\\
DIFF            & Ground Glass Diffuser, 120~Grit (DG10-120; Thorlabs, Inc., Newton, NJ, US)                  		\\
DM1             & SP Dichroic Mirror (001410; Opt Lasers, Warszawa, PL)                                  			\\
DM2             & LP Dichroic Mirror (000424; Opt Lasers, Warszawa, PL)                                  			\\
DM3             & LP Dichroic Mirror, $\lambda_{\mathrm{on}}$~=~613~nm (86-394; Edmund Optics, Barrington, NJ, US)	\\
DM4             & LP Dichroic Mirror, $\lambda_{\mathrm{on}}$~=~552~nm (86-393; Edmund Optics, Barrington, NJ, US)  \\
DM5             & LP Dichroic Mirror, $\lambda_{\mathrm{on}}$~=~480~nm (86-391; Edmund Optics, Barrington, NJ, US)  \\
DM6             & LP Dichroic Mirror, $\lambda_{\mathrm{on}}$~=~427~nm (86-389; Edmund Optics, Barrington, NJ, US)  \\
DPSS-543        & DPSS Laser, 543~nm, 1~W (MGL-FN-543-1W; CNI Laser, Changchun, CN)                      			\\
DPSS-561        & DPSS Laser, 561~nm, 1~W (MGL-FN-561-1W; CNI Laser, Changchun, CN)                      			\\
DPSS-639        & DPSS Laser, 639~nm, 300~mW (MSL-FN-639; CNI Laser, Changchun, CN)                   				\\
FO              & Fiber Oscillator (CNI Laser, Changchun, CN)                                          				\\
HWP1            & Wave Rotator Plate (001832; Opt Lasers, Warszawa, PL)                               				\\
HWP2            & Half-Wave Plate (AHWP05M-580; Thorlabs, Inc., Newton, NJ, US)                               		\\
L1              & Aspheric Lens, \textit{f}~=~4.02~mm (001046; Opt Lasers, Warszawa, PL)                       		\\
L2              & Aspheric Lens, \textit{f}~=~8.00~mm (A240TM-A; Thorlabs, Inc., Newton, NJ, US)                    \\
L3              & Aspheric Lens, \textit{f}~=~4.51~mm (C230TMD-A; Thorlabs, Inc., Newton, NJ, US)                   \\
L4              & Fiber Collimator, \textit{f}~=~10.92~mm (F220SMA-A; Thorlabs, Inc., Newton, NJ, US)               \\
L5              & Aspheric Lens, \textit{f}~=~3.10~mm (C330TMD-A; Thorlabs, Inc., Newton, NJ, US)                   \\
LD-405          & MM Laser Diode, 405~nm, 1~W (HL40033G; Ushio, Inc., Tokyo, JP)                    				\\
LD-450          & MM Laser Diode, 450~nm, 1.6~W (PLTB450B; Osram, Munich, DE)                  						\\
LD-462          & MM Laser Diode, 462~nm, 1.4~W (NDB7675; Nichia Co., Tokushima, JP)                    			\\
LD-525          & MM Laser Diode, 525~nm, 1~W (NDG7475; Nichia Co., Tokushima, JP)                    				\\
LD-637          & MM Laser Diode, 637~nm, 1.2~W (HL63283HD; Ushio, Inc., Tokyo, JP)                 				\\
LD-659          & MM Laser Diode, 659~nm, 1.2~W (HL65213HD; Ushio, Inc., Tokyo, JP)                 				\\
M1              & RGB Mirror (000874; Opt Lasers, Warszawa, PL)                                       				\\
M2              & Broadband Dielectric Mirror (BB1-E02; Thorlabs, Inc., Newton, NJ, US)                       		\\
MF1             & MM Fiber, $\diameter$~=~1~mm, L~=~2 m (M59L02; Thorlabs, Inc., Newton, NJ, US)         			\\
MF2             & MM Fiber, $\diameter$~=~105~\textmu{}m, L~=~5~m (M15L05; Thorlabs, Inc., Newton, NJ, US)   		\\
ND 				& Neutral Density Filter, OD~=~0.8 (NE08A-A; Thorlabs, Inc., Newton, NJ, US) 						\\
PBS             & Polarized Beam Splitter Cube (PBS; Opt Lasers, Warszawa, PL)                        				\\
PD              & Photodiode (OPT101; Texas Instruments, Dallas, TX, US)                                			\\
RM              & Direct Drive Rotation Mount (DDR25; Thorlabs, Inc., Newton, NJ, US)                         		\\
SHTR            & Beam Shutter (SM1SH1; Thorlabs, Inc., Newton, NJ, US)                                       		\\
\bottomrule
\end{tabular}
\footnotesize\center
Abbreviations: Shortpass (SP), Longpass (LP), Multimode (MM) \\
\label{tab:TableS1}
\end{table}

\clearpage
\begin{table}
\caption{Laser diode packages, drivers, and output powers}
\centering
\begin{tabular}{p{2.1cm} p{2.60cm} p{1.90cm} p{4.10cm} p{2.45cm}}
\toprule
\textbf{Key}	& \textbf{Package}	& \textbf{Peak $\lambda$ (nm)}	& \textbf{Driver}		& \textbf{Power (mW)}*	\\
\midrule
LD-405 			& HL40033G 			& 406 							& LPLDD-5A-24V-TP-H 	& 30					\\
LD-450 			& PLTB450B 			& 446 							& LPLDD-5A-24V-TP-H 	& 15					\\
LD-462 			& NDB7675 			& 468 							& LPLDD-5A-24V-TP-H 	& 25					\\
LD-525 			& NDG7475 			& 522 							& LPLDD-5A-24V-TP-H 	& 15					\\
DPSS-543 		& MGL-FN-543 		& 543 							& PSU-H-LED 			& 25					\\
DPSS-561 		& MGL-FN-561 		& 562 							& PSU-H-LED				& 25					\\
LD-637 			& HL63283HD 		& 635 							& LPLDD-5A-24V-TP-H 	& 15					\\
DPSS-639 		& MSL-FN-639 		& 639 							& PSU-H-FDA 			& 50					\\
LD-659 			& HL65213HD 		& 657 							& LPLDD-5A-24V-TP-H 	& 30					\\
\bottomrule
\end{tabular}
\footnotesize\center
*~Output power emitted by a single diode, measured at the distal tip of the colonoscope \\
\label{tab:TableS2}
\end{table}

\clearpage
\begin{table}
\caption{\textit{In~vivo} colonoscopy study data summary}
\centering
\begin{tabular}{lllccccc}
\toprule
\textbf{Patient \#}	& \textbf{Sample \#}	& \textbf{Histology}	& \textbf{WLE}	& \textbf{NBI}	& \textbf{PSE}	& \textbf{LSCI}	& \textbf{Multispectral}	\\
\midrule																																						
P1					& S1 					& Tubular				& \checkmark	& 				& \checkmark	& 				&							\\
P2					& S1 					& Serrated				& \checkmark	& 				& \checkmark	& 				&							\\
                    & S2 					& Serrated				& \checkmark	& 				& \checkmark	& 				&							\\
P3					& S1 					& Serrated				& \checkmark	& 				& \checkmark	& 	            &							\\
P4					& S1 					& Tubular				& \checkmark	& 				& \checkmark	& \checkmark	&							\\
                    & S2 					& Tubular				& \checkmark	& 				& \checkmark	& 	            &							\\
                    & S3 					& Tubular				& \checkmark	& \checkmark	& \checkmark	& \checkmark	&							\\
P5					& S1 					& Tubular				& \checkmark	& 				& \checkmark	& \checkmark	&							\\
P6					& S1 					& Tubular				& \checkmark	& 				& \checkmark	& \checkmark	&							\\
P7					& S1 					& Serrated				& \checkmark	& \checkmark	& \checkmark	& 				&							\\
P8					& S1 					& Serrated				& \checkmark	& \checkmark	& \checkmark	&	            &							\\
P9					& S1 					& Tubular				& \checkmark	& 				& \checkmark	& \checkmark	&							\\
P10					& S1 					& Tubular				& \checkmark	& 				& \checkmark	& \checkmark	&							\\
P11					& S1 					& Tubulovillous			& \checkmark	& \checkmark	& \checkmark	& \checkmark	&							\\
                    & S2 					& Tubular				& \checkmark	& 				& \checkmark	& \checkmark	&							\\
P12					& S1 					& Tubular				& \checkmark	& \checkmark	& \checkmark	& 				&							\\
P13					& S1 					& Tubular				& \checkmark	& \checkmark	& \checkmark	& \checkmark	&							\\
                    & S2 					& Tubular				& \checkmark	& 				& \checkmark	& 				&							\\
P14					& S1 					& Serrated				& \checkmark	& \checkmark	& \checkmark	& \checkmark	& \checkmark				\\
                    & S2 					& Serrated				& \checkmark	& \checkmark	& \checkmark	& 				& \checkmark				\\
P15					& S1 					& Tubular				& \checkmark	& \checkmark	& \checkmark	& \checkmark	& 							\\
                    & S2 					& Tubular				& \checkmark	& \checkmark	& \checkmark	& \checkmark	& 							\\
P16					& S1 					& Tubular				& \checkmark	& \checkmark	& \checkmark	&				& 							\\
                    & S2 					& Tubular				& \checkmark	& \checkmark	& \checkmark	& \checkmark	& \checkmark				\\
P17					& S1 					& Tubular				& \checkmark	& \checkmark	& \checkmark	& \checkmark	& \checkmark				\\
P18					& S1 					& Tubular				& \checkmark	& \checkmark	& \checkmark	& \checkmark	& \checkmark				\\
                    & S2 					& Tubular				& \checkmark	& \checkmark	& \checkmark	& \checkmark	& \checkmark				\\
P19					& S1 					& Tubular				& \checkmark	& \checkmark	& \checkmark	& \checkmark	& \checkmark				\\
                    & S2 					& Tubular				& \checkmark	& \checkmark	& \checkmark	& \checkmark	& \checkmark				\\
P20					& S1 					& Tubular				& \checkmark	& \checkmark	& \checkmark	& \checkmark	& \checkmark				\\
                    & S2 					& Tubular				& \checkmark	& \checkmark	& \checkmark	& \checkmark	& \checkmark				\\
\bottomrule
\end{tabular}
\label{tab:TableS3}
\end{table}

\clearpage
\begin{table}
\centering
\caption{Paired functional comparisons between lesion and normal tissue}
\begin{threeparttable}
\footnotesize
\setlength{\tabcolsep}{10pt}
\begin{tabular}{
l
S[table-format=2.0]
S[table-format=-1.3]
@{\hspace{6pt}} @{}l@{}
S[table-format=-1.4]
@{,\ } S[table-format=-1.4]
@{}l@{} @{\hspace{6pt}}
S[table-format=1.4]
S[table-format=1.4]
S[table-format=-1.4]
}
\toprule
\bfseries Measurement 
& {\bfseries $\boldsymbol{N}$} 
& {\bfseries $\boldsymbol{\Delta\mu}$}
& \multicolumn{4}{c}{\bfseries 95\% CI}
& {\bfseries $\boldsymbol{p_{\scriptscriptstyle\mathrm{sw}}}$}
& {\bfseries $\boldsymbol{p}$}
& {\bfseries $\boldsymbol{d_z}$} \\
\midrule

$\mathrm{StO_2}$ (\%)   & 10    &  2.112    & [ & -0.2302  &  4.4535  & ]   & 0.6115    &           0.0718  &  0.6450 \\
Flow                    & 20    & -0.029    & [ & -0.0566  & -0.0006  & ]   & 0.9605    & \bfseries 0.0456  & -0.4784 \\
\bottomrule
\end{tabular}
\vspace{8pt}
\begin{tablenotes}[flushleft]\footnotesize
\item $\Delta\mu$: Paired mean difference (lesion $-$ normal)
\item 95\% CI: Confidence interval of the paired mean difference
\item $p_{\scriptscriptstyle\mathrm{sw}}$: Shapiro--Wilk $p$-value for normality of paired differences
\item $p$: Two-sided paired $t$-test $p$-value
\item $d_z$: Standardized paired effect size (Cohen’s)
\end{tablenotes}
\end{threeparttable}
\label{tab:TableS4}
\end{table}

\clearpage
\begin{table}
\centering
\caption{Paired CIEDE color difference comparisons between modalities}
\begin{threeparttable}
\footnotesize
\setlength{\tabcolsep}{3pt}
\begin{tabular}{
l l S[table-format=2.0] S[table-format=-2.3]
@{\hspace{4pt}} @{}l@{} S[table-format=-2.3] @{,\ } S[table-format=-2.3] @{}l@{} @{\hspace{4pt}}
@{\hspace{6pt}} S[table-format=1.3]
@{\hspace{6pt}} l @{\hspace{6pt}}
@{\hspace{4pt}} S[table-format=1.3e-2]
S[table-format=-1.3]
}
\toprule
\bfseries Baseline & \bfseries Comparison
& {\bfseries $\boldsymbol{N}$}
& {\bfseries $\boldsymbol{\Delta\mu}$}
& \multicolumn{4}{c}{\bfseries 95\% CI}
& {\bfseries $\boldsymbol{p_{\scriptscriptstyle\mathrm{sw}}}$}
& \bfseries Test
& {\bfseries $\boldsymbol{p}$}
& {\bfseries $\boldsymbol{d_z}$} \\
\midrule
\multirow{2}{*}{Sim-WLE}
& Sim-NBI   & 10    & -0.585    & [ & -1.187 &  -0.145   & ]    & 0.006 & Wilcoxon          & \bfseries 3.711e-02   & -0.645  \\
& Sim-SE    & 10    &  13.618   & [ & 11.584   & 15.653  & ]    & 0.328 & Paired $t$-test   & \bfseries 1.040e-07   &  4.788  \\
\bottomrule
\end{tabular}
\vspace{8pt}
\begin{tablenotes}[flushleft]
\footnotesize
\item $\Delta\mu$: Paired mean difference (comparison $-$ baseline)
\item 95\% CI: Confidence interval of the paired mean difference
\item $p_{\scriptscriptstyle\mathrm{sw}}$: Shapiro--Wilk $p$-value for normality of paired differences
\item Test: Statistical test applied (Wilcoxon signed-rank or two-sided paired $t$-test)
\item $p_{\scriptscriptstyle\mathrm{raw}}$: $p$-value
\item $d_z$: Standardized paired effect size (Cohen’s)
\end{tablenotes}
\end{threeparttable}
\label{tab:TableS5}
\end{table}

\clearpage
\begin{table}
\centering
\caption{Paired RMS contrast comparisons between modalities}
\begin{threeparttable}
\footnotesize
\setlength{\tabcolsep}{3pt}
\begin{tabular}{
l
l
S[table-format=2.0]
S[table-format=-2.5]
@{\hspace{4pt}} @{}l@{}
S[table-format=-2.3]
@{,\ } S[table-format=-2.3]
@{}l@{} @{\hspace{4pt}}
@{\hspace{6pt}} S[table-format=1.3]
@{\hspace{6pt}} l @{\hspace{6pt}}
@{\hspace{4pt}} S[table-format=1.3]
S[table-format=1.3]
S[table-format=-1.3]
}
\toprule
\bfseries Baseline
& \bfseries Comparison
& {\boldmath$N$}
& {\boldmath$\Delta\mu$}
& \multicolumn{4}{c}{\bfseries 95\% CI}
& {\boldmath$p_{\scriptscriptstyle\mathrm{sw}}$}
& \bfseries Test
& {\boldmath$p_{\scriptscriptstyle\mathrm{raw}}$}
& {\boldmath$p_{\scriptscriptstyle\mathrm{adj}}$}
& {\boldmath$d_z$} \\
\midrule

\multirow{7}{*}{WLE Hue}
& NBI Hue           & 19    & -0.00846  & [ & -0.035 &  0.018 & ]   & 0.062 & Paired $t$-test   & 0.516 &           1.000   & -0.152 \\
& NBI Sat.          & 19    &  0.00555  & [ & -0.020 &  0.031 & ]   & 0.275 & Paired $t$-test   & 0.653 &           1.000   &  0.105 \\
& SE Hue            & 10    &  0.15925  & [ &  0.082 &  0.237 & ]   & 0.987 & Paired $t$-test   & 0.001 & \bfseries 0.008   &  1.474 \\
& SE Sat.           & 10    &  0.04337  & [ &  0.005 &  0.082 & ]   & 0.223 & Paired $t$-test   & 0.032 &           0.128   &  0.801 \\
& Height            & 31    &  0.04903  & [ &  0.020 &  0.078 & ]   & 0.565 & Paired $t$-test   & 0.001 & \bfseries 0.009   &  0.629 \\
& Flow              & 20    &  0.02489  & [ &  0.008 &  0.041 & ]   & 0.977 & Paired $t$-test   & 0.005 & \bfseries 0.025   &  0.708 \\
& $\mathrm{StO_2}$  & 10    &  0.02059  & [ & -0.002 &  0.043 & ]   & 0.492 & Paired $t$-test   & 0.072 &           0.216   &  0.645 \\
\midrule
\multirow{7}{*}{WLE Sat.}
& NBI Hue           & 19    & -0.02340  & [ & -0.045 & -0.001 & ]   & 0.042 & Wilcoxon          & 0.018 &           0.088   & -0.465 \\
& NBI Sat.          & 19    & -0.00939  & [ & -0.026 &  0.007 & ]   & 0.753 & Paired $t$-test   & 0.247 &           0.969   & -0.274 \\
& SE Hue            & 10    &  0.13863  & [ &  0.061 &  0.216 & ]   & 0.846 & Paired $t$-test   & 0.003 & \bfseries 0.017   &  1.283 \\
& SE Sat.           & 10    &  0.02275  & [ & -0.018 &  0.064 & ]   & 0.399 & Paired $t$-test   & 0.242 &           0.969   &  0.396 \\
& Height            & 31    &  0.04040  & [ &  0.018 &  0.063 & ]   & 0.919 & Paired $t$-test   & 0.001 & \bfseries 0.006   &  0.670 \\
& Flow              & 20    &  0.00724  & [ & -0.013 &  0.027 & ]   & 0.938 & Paired $t$-test   & 0.463 &           0.969   &  0.168 \\
& $\mathrm{StO_2}$  & 10    & -0.00003  & [ & -0.019 &  0.023 & ]   & 0.013 & Wilcoxon          & 0.557 &           0.969   & -0.001 \\

\bottomrule
\end{tabular}
\vspace{8pt}
\begin{tablenotes}[flushleft]
\footnotesize
\item $\Delta\mu$: Paired mean difference (comparison $-$ baseline)
\item 95\% CI: Confidence interval of the paired mean difference
\item $p_{\scriptscriptstyle\mathrm{sw}}$: Shapiro--Wilk $p$-value for normality of paired differences
\item Test: Statistical test applied (Wilcoxon signed-rank or two-sided paired $t$-test)
\item $p_{\scriptscriptstyle\mathrm{raw}}$: Unadjusted $p$-value
\item $p_{\mathrm{adj}}$: Holm–Bonferroni–adjusted $p$ value (WLE-Hue/WLE-Sat families; values capped at 1)
\item $d_z$: Standardized paired effect size (Cohen’s)
\end{tablenotes}
\end{threeparttable}
\label{tab:TableS6}
\end{table}

\end{document}